\documentclass[letterpaper,twocolumn,10pt]{article}
\usepackage{usenix-2020-09}

\usepackage[ruled, lined, linesnumbered, commentsnumbered, noend, longend]{algorithm2e}

\usepackage{xspace}
\usepackage{tikz}
\usepackage{amsmath}

\usepackage{pgfplots}
\pgfplotsset{compat=1.16}

\usepackage{subcaption}

\usepackage{xcolor}
\usepackage{booktabs}
\usepackage{url}
\usepackage[subtle]{savetrees}

\usepackage{makecell}
\usepackage{array}

\usepackage[capitalise,nameinlink]{cleveref}
\crefname{algorithm}{Algorithm}{Algorithms}
\Crefname{algorithm}{Algorithm}{Algorithms}
\crefname{section}{\S\!}{\S\S\!}
\Crefname{section}{Section}{Sections}
\crefname{figure}{Figure}{Figures}
\Crefname{figure}{Figure}{Figures}
\crefname{equation}{Equation}{Equations}
\Crefname{equation}{Equation}{Equations}
\crefname{listing}{Listing}{Listings}
\Crefname{listing}{Listing}{Listings}
\crefname{defn}{definition}{definitions}

\usepackage[inline]{enumitem}
\newlist{inlist}{enumerate*}{1}
\setlist[inlist]{label=\textbf{(\arabic*)}}

\usepackage{authblk}

\newcommand{\mashup}{\textsc{MashUp}\xspace}
\newcommand{\bsic}{\textsc{BSIC}\xspace}
\newcommand{\resail}{\textsc{RESAIL}\xspace}
\newcommand{\sail}{\textsc{SAIL}\xspace}
\newcommand{\dxr}{\textsc{DXR}\xspace}
\newcommand{\hibst}{\textsc{Hi-BST}\xspace}
\newcommand{\cram}{\textsc{CRAM}\xspace}

\begin{document}

\date{}

\title{\vspace{-7em}\Large \bf Scaling IP Lookup to Large Databases using the CRAM Lens\vspace{-1em}}

\makeatletter
\renewcommand\AB@affilsepx{\quad\qquad\protect\Affilfont}
\makeatother
\setlength{\affilsep}{0.8em}
\renewcommand\Affilfont{\itshape}
\renewcommand\Authands{\qquad}
\renewcommand\Authsep{\qquad}
\author[1]{Robert Chang}
\author[1]{Pradeep Dogga}
\author[2]{Andy Fingerhut}
\author[1]{Victor Rios}
\author[1]{George Varghese}
\affil[1]{University of California, Los Angeles}
\affil[2]{Cisco Systems}

\maketitle


\begin{abstract}

Wide-area scaling trends require new approaches to Internet Protocol (IP) lookup, enabled by modern networking chips such as Intel Tofino~\cite{intel_tofino}, AMD Pensando~\cite{amd_pensando}, and Nvidia BlueField~\cite{nvidia_bluefield}, which provide substantial ternary content-addressable memory (TCAM) and static random-access memory (SRAM). However, designing and evaluating scalable algorithms for these chips is challenging due to hardware-level constraints. To address this, we introduce the \cram~(CAM+RAM) lens, a framework that combines a formal model for evaluating algorithms on modern network processors with a set of optimization idioms. We demonstrate the effectiveness of \cram by designing and evaluating three new IP lookup schemes: \resail, \bsic, and \mashup. \resail enables Tofino-2 to scale to 2.25 million IPv4 prefixes---likely sufficient for the next decade---while a pure TCAM approach supports only 250k prefixes, just 27\% of the current global IPv4 routing table. Similarly, \bsic scales to 390k IPv6 prefixes on Tofino-2, supporting 3.2 times as many prefixes as a pure TCAM implementation. In contrast, existing state-of-the-art algorithms, \sail~\cite{sail} for IPv4 and \hibst~\cite{hibst} for IPv6, scale to considerably smaller sizes on Tofino-2.

\end{abstract}


\section{Introduction}
\label{sec:introduction}

For many, Internet Protocol~(IP) lookup is considered a challenge of the past. With over 40 years of research and hundreds of papers (e.g., ~\cite{dir, lulea, lampson, dxr, tree_bitmap, dharma, poptrie, hibst, sail}) focused on supporting IP lookup at scale, numerous schemes have been developed---some of which have been in practical use for two decades. However, these classical approaches are {\em single-resource} solutions, designed for conventional switch chip architectures that provided either specialized hardware like ternary content-addressable memory~(TCAM) or commodity random-access memory~(RAM), such as on-chip static RAM~(SRAM) coupled with off-chip dynamic RAM~(DRAM), but not both. TCAM enables parallel searches across wildcarded entries in a single clock cycle but requires three times more transistors per bit than SRAM and consumes hundreds of watts~\cite{tcampower}. RAM-based approaches are cheaper but require additional complexity and memory compared to pure TCAM solutions. Thus far, commercial switch chip vendors have scaled these {\em single-resource} solutions by increasing hardware resources.

In this paper, we contend that it is important to reconsider IP lookup due to the continued growth of lookup tables and a recent inflection point in network hardware. A slew of new application-specific integrated circuits~(ASICs), such as Intel Tofino~\cite{intel_tofino}, AMD Pensando~\cite{amd_pensando}, and Nvidia BlueField~\cite{nvidia_bluefield}, have transformed the networking chip market~\cite{p4_ecosystem}. These chips are built on two modern packet processing architectures---Reconfigurable Match-Action Tables~(RMT)\cite{rmt} and disaggregated RMT~(dRMT)\cite{drmt}---which consist of match-action processors with access to large amounts of {\em both} TCAM and SRAM. We review these architectures at the start of \cref{sec:cram}. This leads us to our central question: \textit{How can we leverage modern networking chips, utilizing both TCAM and SRAM, to develop new IP lookup algorithms that scale to larger databases than classical single-resource solutions?}

Two main challenges make designing scalable algorithms for RMT and dRMT chips difficult:
\begin{inlist}
  \item Lack of an abstract model for evaluating and comparing algorithms. Chip-specific arcana such as memory allocation, metadata storage, and action bits must be carefully considered.
  \item Large but finite resources. While TCAM, SRAM, and pipeline stages are available in substantial amounts, they require careful algorithm design to scale effectively.
\end{inlist}

\textbf{Solution}: We introduce the \cram~(CAM+RAM) lens, an abstract model of modern packet processing architectures, such as RMT and dRMT, paired with a set of optimization idioms. The \cram model enables us to estimate algorithm scalability using higher-order space and time metrics, without requiring simulation of ASIC-specific details such as TCAM block sizes, SRAM page sizes, and per-stage memory. The \cram model goes beyond classical models such as random-access machine~(RAM)~\cite{ram} and parallel RAM~(PRAM)~\cite{pram} by adding TCAM operations and using match-action dependencies to measure time complexity. The optimization idioms provide eight strategies for designing scalable algorithms.

\cram can be generalized to other hardware architectures, such as 
smart network interface cards~(SmartNICs)~\cite{intel_ipu, amd_pensando, nvidia_bluefield} and field-programmable gate arrays~(FPGAs)~\cite{intel_fpga, amd_fpga}, and applied to broader network applications like packet classification~\cite{pkt_class_surv1, pkt_class_surv2, pkt_class_surv3} and in-network machine learning~(ML)~\cite{homunculus, planter, inml_survey}. However, this is not the focus of our paper. For completeness, we briefly discuss these extensions in \cref{subsec:other_architectures} and \cref{subsec:other_applications}. Instead, we concentrate on applying \cram to {\em IP lookup} because of the following observations:

\begin{figure}[!t]
    \centering
    \begin{tikzpicture}
    \large
    \begin{axis}[
        xmin = 3, xmax = 23,
        ymin = 1, ymax = 10,
        xtick={3,5,...,23},
        ytick={1,2,...,10},
        axis y line* = left,
        width=\columnwidth*0.9,
        height=170pt,
        xticklabel={\ifdim\tick pt<10pt 0\fi\pgfmathprintnumber{\tick}},
        xticklabel style={rotate=0,font=\small},
        yticklabel style={rotate=0,font=\small},
        xlabel style={font=\normalsize,at={(axis description cs:.5,-.1)}},
        ylabel style={font=\normalsize,at={(axis description cs:-.07,.5)}},
        legend style={at={(0.575,1)},anchor=north east,font=\small},
        xlabel=Year (2003-2023),
        ylabel=Active IPv4 Entries (1x$10^5$),
        tick pos=left
    ]
    \addplot[line width=1.5pt,mark=none,smooth,color=blue]
        coordinates{
            (3,1.18362)
            (4,1.29296)
            (4.5,1.38674)
            (5.5,1.60471)
            (6.33,1.81032)
            (7.75,2.35660)
            (8.4,2.55837)
            (9.5,3.00500)
            (10.5,3.29240)
            (11.2,3.52301)
            (12,3.96490)
            (13,4.45339)
            (14.5,5.00789)
            (15.1,5.38775)
            (16.5,6.18431)
            (17.5,6.75741)
            (18.5,7.24538)
            (19.9,8.12839)
            (20.9,8.47974)
            (21.9,9.02461)
            (22.4,9.16372)
            (23,9.43586)
        };
    \addlegendentry{AS65000 (IPv4)}
    \addplot[line width=1.5pt,mark=none,densely dashed,color=red]
        coordinates{
            (-100,-100)
        };
    \addlegendentry{AS131072 (IPv6)}
    \end{axis}
    \begin{axis}[
        xmin = 3, xmax = 23,
        ymin = 0, ymax = 20,
        hide x axis,
        hide y axis,
        width=\columnwidth*0.9,
        height=170pt
    ]
    \addplot[line width=1.5pt,mark=none,densely dashed,color=red]
        coordinates{
            (3,0.0405)
            (4,0.05)
            (5,0.0706)
            (6,0.0853)
            (7,0.0793)
            (7.7,0.0918)
            (8,0.113)
            (9,0.1632)
            (10,0.2485)
            (11,0.4158)
            (12,0.7828)
            (13,1.1867)
            (14,1.6179)
            (15,2.0955)
            (16,2.6476)
            (17.1,3.6395)
            (18,4.6645)
            (19,6.3162)
            (20.1,8.0989)
            (21,10.8725)
            (21.9,14.2827)
            (22.4,15.3904)
            (22.8,16.4702)
            (23,18.0344)
        };
    \end{axis}
    \pgfplotsset{every axis y label/.append style={rotate=180}}
    \begin{axis}[
        xmin=0, xmax=1,
        ymin=0, ymax=20,
        hide x axis,
        axis y line*=right,
        ylabel=Active IPv6 Entries (1x$10^4$),
        ytick={0,2,...,20},
        width=\columnwidth*0.9,
        height=170pt,
        yticklabel style={rotate=0,font=\small},
        ylabel style={font=\normalsize,at={(axis description cs:1.08,.5)}}
    ]
    \end{axis}
    \end{tikzpicture}
    \vspace{-1em}
    \caption{BGP routing table size over the past two decades}
    \label{fig:bgp_growth}
    \vspace{-\baselineskip}
\end{figure}
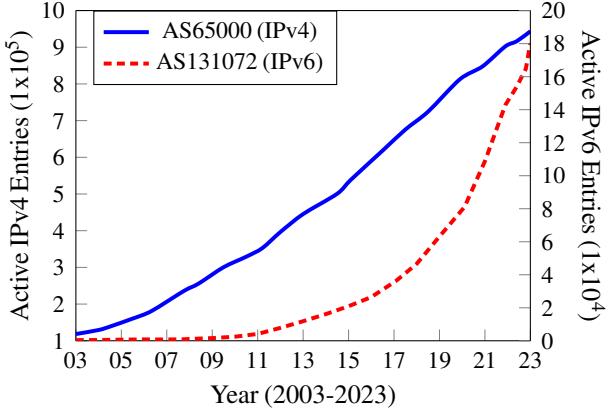

{\bf O1. Continued IPv4 growth:} Over the past two decades, the global IPv4 routing table has grown {\em linearly}~\cite{cidr_report_ipv4, bgp_report_ipv4}, doubling in size every decade (\cref{fig:bgp_growth}). If this trend continues, the IPv4 table could reach two million entries by 2033.

{\bf O2. Rapid IPv6 deployment:} In the same period, the global IPv6 routing table has grown {\em exponentially}~\cite{cidr_report_ipv6, bgp_report_ipv6}, doubling every three years (\cref{fig:bgp_growth}). Even if growth slows to a linear rate, the IPv6 table could still reach half a million entries by 2033. IPv6 prefixes are also four times wider, though typically, only the first 64 bits are used for global routing.

{\bf O3. Virtual private networks~(VPNs):} Some routers maintain hundreds of VPN routing tables. On such devices, publicly available routing tables account for only a fraction of the total capacity required.

{\bf O4. Other tasks:} Routers need table memory for additional tasks such as network address translation~(NAT) and firewalls. Minimizing the memory used for forwarding allows more features to fit on a chip.

\cram enables us to rethink pre-existing IP lookup schemes, such as \sail~\cite{sail}, \dxr~\cite{dxr}, and multibit tries, to develop new algorithms that scale to larger databases. We start with the best-of-breed algorithms from three classic IP lookup approaches---search on prefix {\em lengths}, search on prefix {\em ranges}, and trie-based search. We derive new algorithms using the optimization idioms~(\cref{subsec:optimization_idioms}) and predict their scalability with the \cram model~(\cref{subsec:cram_model}). Each new algorithm---rethinking \sail~(\resail), Binary Search with Initial CAM~(\bsic), and \mashup---offers unique strengths for different settings.

We chose Intel Tofino-2, an RMT switch chip, for our experiments because we had access to its development environment. While Intel recently announced it will not develop new Tofino models, it remains committed to supporting Tofino and Tofino-2~\cite{nick_letter}. We expect our results to hold for dRMT, as RMT is a stricter version of dRMT with additional access restrictions.

This paper makes the following contributions:

\textbf{1. IP Lookup Algorithms:} Three new scalable IP lookup algorithms---\resail, \bsic, and \mashup.

\textbf{2. \cram Model:} An abstract model for quickly estimating the scalability of packet processing algorithms before detailed implementation.

\textbf{3. Optimization Idioms:} Eight design strategies for minimizing TCAM, SRAM, and pipeline stages.

\textbf{4. Evaluation:} Simulations and Tofino-2 implementations achieve 9X (IPv4) and 3X (IPv6) improvements over pure TCAM solutions, enabling scalability for the next decade.

The remainder of this paper is organized as follows. \cref{sec:cram} introduces the \cram model and optimization idioms, followed by an overview of the idioms in action. It also briefly explores other hardware architectures, broader network applications, and algorithmic requirements. We design three new IP lookup algorithms in \cref{sec:resail}, \cref{sec:bsic}, and \cref{sec:mashup}. \cref{sec:results} presents simulation and implementation results, followed by scalability experiments in \cref{sec:scalability}. \cref{sec:crammodelevaluation} evaluates the predictive accuracy of the \cram model, and \cref{sec:related} surveys related work. We conclude in \cref{sec:conclusion}.
\newcommand*\BitAnd{\mathbin{\&}}
\newcommand*\BitOr{\mathbin{|}}
\newcommand*\BitXor{\mathbin{\hat{}}}
\newcommand*\ShiftLeft{<<}
\newcommand*\ShiftRight{>>}
\newcommand*\BitNeg{\ensuremath{\mathord{\sim}}}


\section{The \cram Lens}
\label{sec:cram}

We now formally introduce the \cram model, list eight optimization idioms, and preview the idioms in action. Additionally, we briefly discuss how \cram can be generalized to other hardware architectures, applied to broader network applications, and adapted to algorithmic requirements.

The \cram model abstracts two modern packet processing architectures, RMT~\cite{rmt} and dRMT~\cite{drmt}. Figures depicting both can be found in \cref{subsec:rmt_drmt_overview}. A list of known RMT and dRMT implementations is provided in \cref{subsec:rmt_drmt_implementations}.

{\bf RMT}: RMT is a sequential pipeline architecture of match-action stages. TCAM and SRAM are partitioned among stages such that a stage cannot access the memory of other stages.

{\bf dRMT}: In contrast, dRMT features programmable processors that execute match-action operations in any order. It disaggregates memory from processors by relocating TCAM and SRAM into a shared external pool.

\subsection{The \cram Model} \label{subsec:cram_model}

The \cram model adds two extensions to the RAM model~\cite{ram}: first, the ability to perform an SRAM {\em or} TCAM table lookup; second, an explicit dependency structure between steps (as in RMT compilers~\cite{rmt_compilers}) that models the ability to execute multiple steps in parallel. Our goal is for the memory and run time measures of a \cram model program to be within a small constant factor of the measures for actual hardware implementations.

Thus, a \cram model program is parameterized by:

\begin{itemize}
  \item A register size $w$, and a set $R$ of ($w$-bit) registers. Let $C$ denote the set of $w$-bit integers in the range $[0, 2^w-1]$.
  \item Sets of unary ($Uops$) and binary ($Bops$) operators on $w$-bit values, e.g., $Uops = \{ +, -, \BitNeg, ! \}$ and $Bops = \{ +, -, \ShiftLeft, \ShiftRight, ==, \mathrel{\mathtt{!=}}, <, \leq, >, \geq, \BitAnd, \BitOr, \BitXor, \&\&, || \}$, with behavior as defined in languages like Java and P4~\cite{p4_ecosystem}.
\end{itemize}

A \cram model program consists of a parser $P$, a deparser $D$, and a directed acyclic graph $G$ comprised of {\em steps}. A state $S$ is a function from $R$ to $C$. $P$ is a function from all bit sequences representing packets to an initial state. $D$ is a function from a final state back to all bit sequences representing packets.

A {\em step} may optionally begin with a single table lookup operation. A table $t$ consists of a match kind (exact or ternary), a key selector function $K_t$, a maximum number of entries $n_t$, and a default value $Z_t$. $K_t$'s result is a sequence of $k_t$ bits, each representing a chosen bit position within one register of $R$. An entry $e$ contains a key and associated data, with the $d_t$ bits of associated data stored in a set of $w$-bit registers $A_t$.

For an exact match table, the key is a $k_t$-bit integer. A special case arises for exact match tables with $n_t = 2^{k_t}$, in which the key does not need to be explicitly stored, as it can be used to directly index into the table. For a ternary match table, the key is a pair of $k_t$-bit integers, a value $v_e$ and a mask $m_e$, plus an integer priority $p_e$. All keys in the same table must be distinct.

A step consists of an optional table $t$ followed by a sequence of statements in the form $if (cond): dest = expr$. Here, $dest$ is an element of $R$, $expr$ contains a single unary or binary operator with operands from $R \cup A_t \cup C$, and $cond$ is a potentially nested expression with operands from $R \cup A_t \cup C$. No data dependencies are allowed within this sequence, i.e., for any statement in the sequence that assigns a value to $r \in R$, $r$ may not appear in $cond$ or $expr$ of any later statements. This enables all statements within a step to be executed in parallel.

A step reads register $r$ if any bit of $r$ appears in the output of its key selector function $K_t$, or as part of $cond$ or $expr$ in any of its statements. A step writes register $r$ if $r$ appears as $dest$ in any of its statements.

For all steps $u$ and $v$ in $G$, if $u$ writes $r$ and $v$ reads or writes $r$, then there must be a directed path $(u, v)$ or $(v, u)$. This prevents $u$ and $v$ from being executed in parallel. This condition must hold for all registers. A directed path $(u, v)$ indicates that step $u$ must be executed before step $v$. If there is no directed path between two steps, they may be executed in parallel.

The \cram model introduces a set of higher-order space and time metrics. The memory footprint of a \cram model program is evaluated by calculating the total TCAM and SRAM bits across all tables $t$ in $G$. In a ternary (exact) match table, the memory used for the keys is $n_t k_t$ TCAM (SRAM) bits. For ternary match tables, we only count the $v_e$ component of the key, as these are the logical bits involved in the match. For both types of tables, the memory used for the associated data is $n_t d_t$ SRAM bits. To compute the overall TCAM and SRAM totals, simply sum the bits used across all tables. The latency of a \cram model program is evaluated by determining the number of steps (nodes) in the longest directed path in $G$.

\subsection{Optimization Idioms}\label{subsec:optimization_idioms}

The following idioms can be applied together in various combinations to achieve different space-time trade-offs:

{\em I1. Compress with TCAM}: Entries containing wildcards must be expanded to fit into SRAM. For example, the prefix 1** would be stored as 100, 101, 110, and 111. However, by utilizing TCAM, these four SRAM entries can be compressed into a single TCAM entry (1**), thus saving nine bits.

{\em I2. Expand to SRAM:} In the dual of {\em I1}, replace a TCAM block with SRAM if the expanded forms of its prefixes are less than a small constant factor $c$ of the original TCAM entries. We choose $c=3$ because TCAM requires three times more transistors per bit than SRAM~\cite{transistors}.

{\em I3. Compress with SRAM:} Despite their high memory cost, directly indexed data structures such as next hop arrays are used because they avoid the extra instructions needed for hashing. However, since most RMT and dRMT ASICs are designed with the cost of performing SRAM-based lookups---whether by hashing or direct indexing---being exactly the same, it is often more advantageous to use compressed forms of SRAM storage such as hash tables instead.

{\em I4. Strategic Cutting:} If several entries at a given node share a common prefix, we can save memory by strategically cutting at the bit position where the shared prefix ends, storing only one copy of the repeated bits. While this is how multibit tries~\cite{prefix_expansion} work, we extend the concept to TCAM nodes.

{\em I5. Table Coalescing:} To reduce memory waste, minimally populated logical tables can be coalesced in shared physical TCAM blocks or SRAM pages. They can be differentiated with tag bits~\cite{tag_bits}. Although tagging increases the lookup key width, it minimizes physical TCAM and SRAM fragmentation.

{\em I6. Look-aside TCAM:} IP lookup schemes are often optimized around common cases such as 24-bit IPv4 prefixes. As a result, uncommon entries (e.g., extremely short or long prefixes) tend to require undue computational or storage costs. We address this by moving the special prefixes into a separate look-aside TCAM that can be trivially searched in parallel.

{\em I7. Step Reduction:} A program's number of steps can be reduced by leveraging match-action unit (MAU) parallelism to consolidate data-independent lookups into a single stage.

{\em I8. Memory Fan-out:} In traditional RAM model architectures, a lookup table can be accessed multiple times per packet. However, many RMT and dRMT chips restrict each table to one memory access per packet. To address this limitation, we split the original table by fanning out its contents and storing entries accessed by different lookups in separate tables.


\subsection{Idioms in Action}\label{subsec:idioms_in_action}

\begin{figure}[t]
    \centering
    \includegraphics[width=0.47\textwidth]{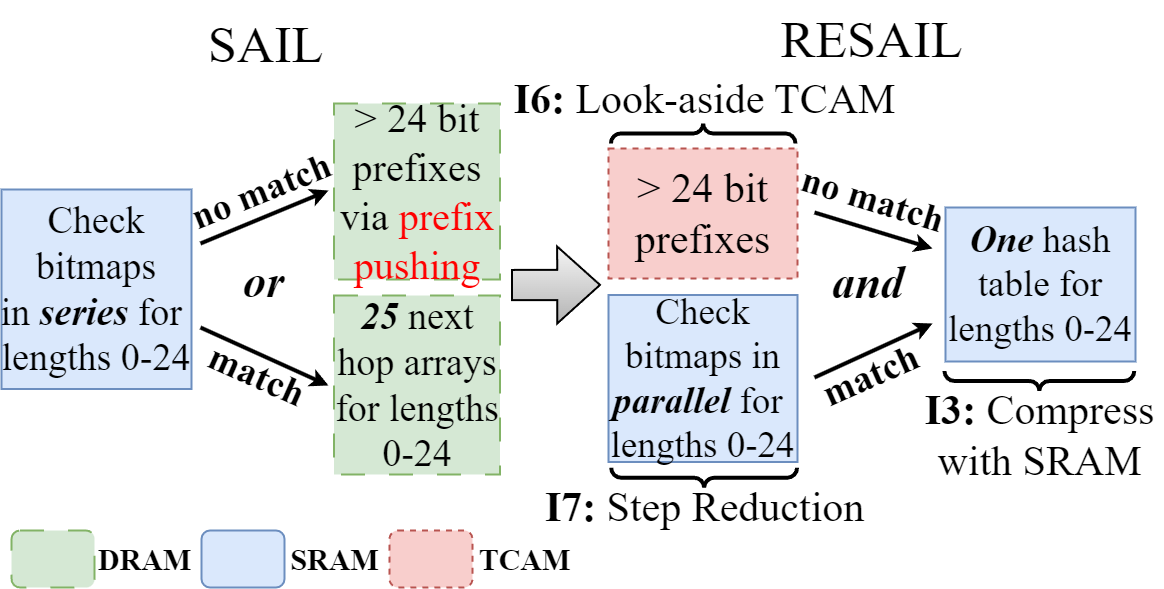}
    \caption{From SAIL to \resail via \cram idioms}
    \label{fig:sailtoresail}
    \vspace{-\baselineskip}
    \vspace{0.75em}
\end{figure}

\begin{figure}[t]
    \centering
    \includegraphics[width=0.45\textwidth]{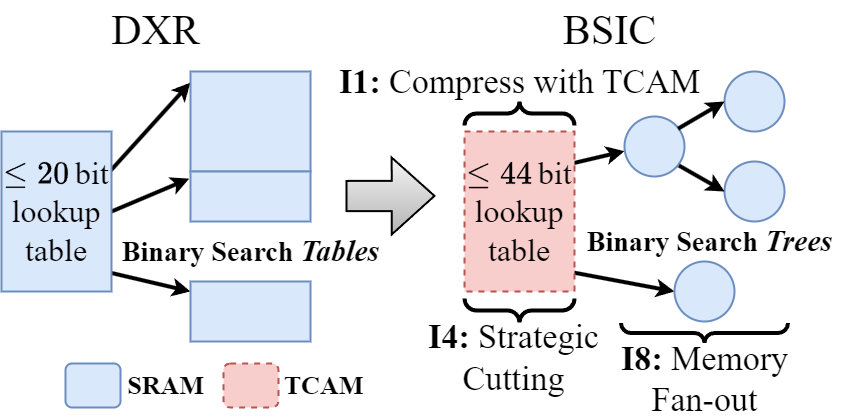}
    \caption{From DXR to BSIC via \cram idioms}
    \label{fig:dxrtobsic}
    \vspace{-\baselineskip}
    \vspace{0.75em}
\end{figure}

\begin{figure}[t]
    \centering
    \includegraphics[width=0.45\textwidth]{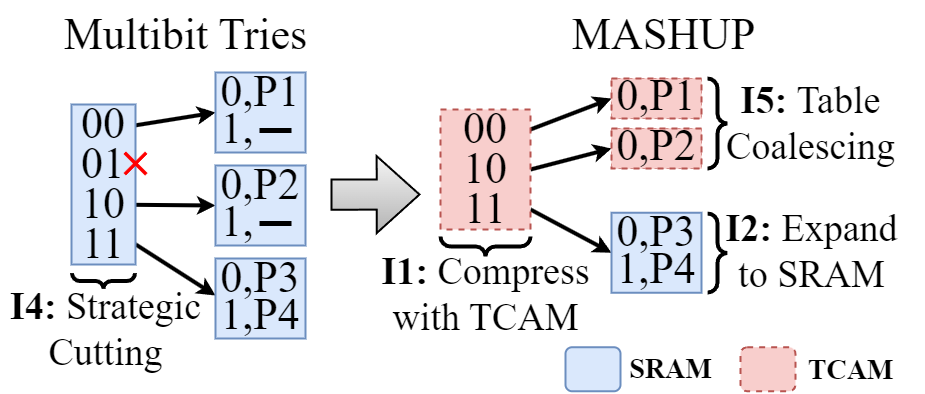}
    \caption{From multibit tries to \mashup via \cram idioms}
    \label{fig:multibittomashup}
    \vspace{-\baselineskip}
\end{figure}

We briefly preview the idioms in action for three fundamental classes of IP lookup: search on prefix {\em lengths}, search on prefix {\em ranges}, and trie-based search.

\textbf{From \sail to \resail:} Search on prefix {\em lengths}~\cite{waldvogel, dharma, sail, binary_search_on_length, dleft} splits IP lookup into two sub-problems: finding the {\em length} of the longest match and retrieving the {\em next hop}. \sail~\cite{sail}, the best performing IPv4 lookup scheme in hardware settings with fast on-chip SRAM and cheap off-chip DRAM, uses a bitmap of length $2^L$ to determine whether there is a matching prefix of length $L$. This works well for prefixes of up to length 24 (the vast majority); for prefixes longer than 24 bits, \sail uses a complex scheme called pivot pushing that requires excessive prefix expansion~\cite{prefix_expansion}. In \cref{fig:sailtoresail}, to obtain \resail~(rethinking \sail), we: {\em I6)} Move the small number of prefixes longer than 24 bits into a separate look-aside TCAM. {\em I3)} Compress all the next hop arrays into a single hash table using a standard encoding trick~\cite{network_algorithmics}. {\em I7)} Use MAU (stage) parallelism to reduce the number of steps by performing the bitmap lookups in parallel.

\textbf{From \dxr to \bsic}: Search on prefix {\em ranges}~\cite{dxr, lampson, dir, suri, changhoon} represents prefixes as range endpoint pairs (e.g., 0**$\rightarrow$[000, 011]). Finding the longest matching prefix becomes equivalent to finding the smallest range that encompasses the lookup key. \dxr~\cite{dxr}, the fastest IPv4 software implementation of range-based searches, uses an initial lookup table to split the search space into multiple smaller binary search tables. In \cref{fig:dxrtobsic}, to obtain \bsic~(Binary Search with Initial CAM), we: {\em I1)} Replace the SRAM-based initial lookup table, which supports up to 20-bit prefixes due to direct indexing, with a TCAM-based table that can store prefixes of up to 44 bits (Tofino-2 TCAM block width). {\em I8)} Replace each binary search {\em table} with a binary search {\em tree} that can be fanned out across stages. {\em I4)} Strategically cut the initial lookup table to balance TCAM required against binary search depth.

\textbf{From multibit tries to \mashup:} For trie-based search~\cite{prefix_expansion, lctrie, lulea, indira, flashtrie, tree_bitmap, poptrie}, we specifically focus on multibit tries---tries that examine multiple bits, known as a stride, per lookup. We do not consider state-of-the-art {\em compressed} trie schemes like Poptrie~\cite{poptrie} and Tree Bitmap~\cite{tree_bitmap}, because in the \cram model, one can directly compress with TCAM without the extra computational and storage costs of bitmap compression. ~\cref{fig:multibittomashup} shows a standard multibit trie for the prefixes P1 = 000*, P2 = 100*, P3 = 110*, and P4 = 111*, with a 2-bit stride at the root and a 1-bit stride at the next level---chosen by strategic cutting {\em (I4)} to minimize the number of downstream pointers. In \cref{fig:multibittomashup}, to obtain \mashup~(mashup of CAM and RAM nodes), we: {\em I1)} Replace the SRAM root node with a TCAM node to eliminate the empty 01 entry, and do the same for the two upper-right nodes. {\em I2)} Leave the bottom-right node as SRAM, as it has no wasted space. {\em I5)} Coalesce the two upper-right TCAM nodes using tag bits~(not shown). While the improvement is minimal in this simple example, \cref{subsec:designing_mashup} demonstrates significant gains for large databases.

Although the \cram versions of these classical schemes may seem simple, they require new algorithms to determine, for example, where to make strategic cuts and which nodes to coalesce. We elaborate on these details when describing \resail, \bsic, and \mashup in \cref{sec:resail}, \cref{sec:bsic}, and \cref{sec:mashup}, respectively.

\subsection{\cram for other Hardware Architectures}\label{subsec:other_architectures}

\cram can be generalized to hardware architectures beyond RMT and dRMT as follows: the space and time metrics of an algorithm specified in the \cram model serve as lower bounds on the corresponding costs in any implementation, whether in programmable switch ASICs~\cite{intel_tofino, cisco_silicon_one, nvidia_spectrum, marvell_teralynx}, SmartNICs~\cite{intel_ipu, amd_pensando, nvidia_bluefield}, FPGAs~\cite{intel_fpga, amd_fpga}, or purpose-built fixed-function ASICs designed solely to execute that algorithm. A faithful implementation of a \cram algorithm achieves a minimum latency equal to that of the longest (critical) path in its directed acyclic graph. While an implementation may have a longer latency, it cannot be shorter. Similarly, the number of bits required may match or exceed the amount specified by the \cram model, but it cannot be less.

\subsection{\cram for broader Network Applications}\label{subsec:other_applications}

Although this paper focuses on IP lookup, we believe \cram applies to other memory-intensive network applications. These include packet classification~(with Access Control Lists~(ACLs) and Quality of Service~(QoS) as specific instances), measurement algorithms~(such as sketching), regular expression matching, and in-network ML.


In packet classification~\cite{pkt_class_surv1, pkt_class_surv2, pkt_class_surv3}, packet headers are matched against a classifier, where the highest-priority match determines whether to allow or deny traffic, enforce a QoS policy, or apply a custom action. Measurement algorithms~\cite{opensketch, hashpipe, count_min_sketch, sketch_sampling}, by contrast, dynamically build and update a stateful database that tracks network statistics such as per-flow counters, traffic volume, or frequency estimates. Regular expression matching~\cite{insomnia, estan, hyperscan} compares unstructured data streams against a database of predefined patterns, often represented as finite automata~\cite{p4dfa}, to detect keywords, signatures, and anomalies. Lastly, in-network ML~\cite{homunculus, planter, inml_survey} performs inference by matching a feature vector against a classification model database, which contains decision rules mapping extracted feature values to inference labels.

In practice, these applications often rely on combinations of common data structures, such as decision trees, Bloom filters~\cite{bloom_filters}, tries, hash tables, and bitmaps, many of which we demonstrate how to optimize with \cram. Consequently, our optimization idioms naturally extend beyond IP lookup. For example, the careful balancing of TCAM compression ({\em I1}) and SRAM expansion ({\em I2}) used in \mashup to create a hybrid trie, can similarly be applied to packet classification~\cite{neurocuts, hypercuts} and in-network ML~\cite{mousika, decision_tree_p4} algorithms that rely on decision trees. Likewise, the look-aside TCAM ({\em I6}) in \resail, which captures longer prefixes, can serve a similar role in offloading other specialized cases, such as multi-field wildcard classification rules~\cite{pkt_class_surv1}, heavy-hitter flows with rare protocols~\cite{hashpipe}, multi-line attack patterns~\cite{hyperscan}, and fast-patch updates for classification models~\cite{IIsy}. A more in-depth exploration of \cram's broader applicability is left for future work.

\subsection{Other Algorithmic Requirements}\label{subsec:constraints}

Certain algorithmic requirements introduce additional considerations. We briefly examine how \cram applies to algorithms that require atomic memory updates, stateful data plane operations, and pseudo-random keys.

\textbf{Atomic memory updates:} \cram neither facilitates nor hinders an algorithm’s ability to support atomic updates~\cite{p4dfa, flowradar, atomic_updates_data_plane}. The \cram model provides an abstraction to estimate {\em performance}, not an execution model to predict {\em implementation feasibility}. If an algorithm requiring atomic updates can be implemented, the \cram model should accurately predict latency and memory.


\textbf{Stateful data plane operations:} P4 register arrays are the primary mechanism for stateful operations in the data plane, as used in \cite{hashpipe, snap, conquest}. Stateful operations can be incorporated into the \cram model by introducing a new SRAM-based register match table, and counting these memory bits separately alongside regular TCAM and SRAM bits.



\textbf{Pseudo-random keys:} For algorithms with pseudo-random keys~\cite{halfsiphash, netcache, aes}, the efficacy of some \cram optimization idioms is clearly reduced, as uniformly distributed random bytes are difficult, if not impossible, to compress. In contrast, other idioms are unaffected by key distribution. Specifically, table coalescing ({\em I5}), look-aside TCAM ({\em I6}), step reduction ({\em I7}), and memory fan-out ({\em I8}) can still be applied. Nevertheless, the benefits of the \cram model are likely to be significantly diminished, as lack of compression eliminates a major source of optimization.
\section{\resail}
\label{sec:resail}

\begin{figure}[t]
     \centering
     \begin{subfigure}[b]{0.49\textwidth}
         \centering
         \includegraphics[width=0.95\textwidth]{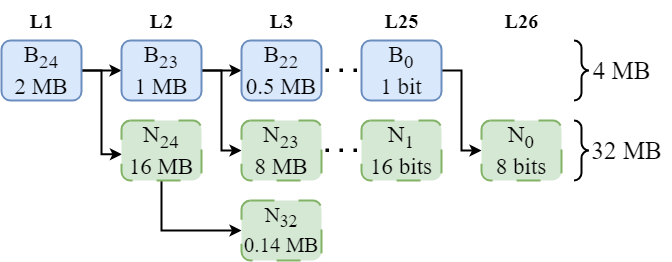}
         \caption{\cram model representation of \sail}
         \label{fig:sail}
     \end{subfigure}
     \begin{subfigure}[b]{0.49\textwidth}
         \centering
         \includegraphics[width=0.75\textwidth]{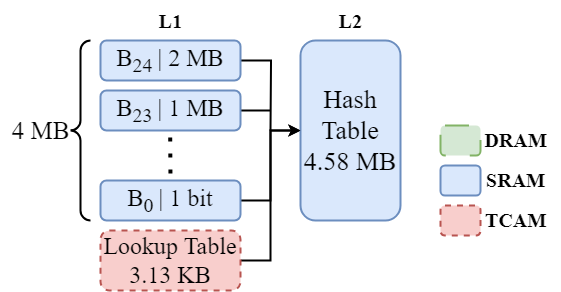}
         \caption{\cram model representation of \resail ($min\_bmp$=0)}
         \label{fig:resail}
     \end{subfigure}
    \caption{\sail vs \resail for IPv4 prefixes in AS65000}
    \label{fig:sailvsresail}
    \vspace{-\baselineskip}
\end{figure}

By applying the optimization idioms to \sail~\cite{sail}, we create a new IPv4 \cram lookup algorithm called \resail. Refer to ~\cref{fig:sailtoresail} for the intuition to which we now add details.

{\bf \sail Review:} \sail designates 24 as a pivot level and divides the forwarding information base~(FIB) into short prefixes~($\leq$24) and long prefixes~(>24). \sail determines whether there is a length-$i$ match for $i \leq 24$ by consulting a bitmap $B_i$ of size $2^{i}$ in which bit $p$ is set if and only if $p$ is a length-$i$ prefix in the FIB. Since the total memory footprint of the bitmaps is 4 MB, they are stored in on-chip SRAM. If a match is found in $B_i$, the next hop is retrieved by directly indexing into a next hop array $N_i$ of size $2^{i}$ located in off-chip DRAM. \sail handles prefixes of length $i > 24$ by using a complex scheme called pivot pushing that expands~\cite{prefix_expansion} them to 32-bit entries in $N_{32}$.

\subsection{Applying the Optimization Idioms} \label{subsec:designing_resail}

We show the \cram derivation of \resail using the IPv4 AS65000 BGP routing table (Sep 2023). Start with the \cram model representation of \sail in \cref{fig:sail}. To obtain \resail in \cref{fig:resail}, use the idioms as follows:

{\bf 1.} Observe a large number~(26) of data dependencies between the bitmaps and next hop arrays. This makes sense in the RAM model as it enables early exits which reduce average execution time. However, these are false dependencies because their lookup keys can be computed in parallel. Therefore, we apply step reduction~({\em I7}) to reduce all the bitmap and next hop array lookups into a single step. The next hop can then be determined by taking the highest priority result.

{\bf 2.} \sail relies on a significant amount of DRAM~(32 MB) for its directly indexed next hop arrays. Since DRAM is not available in \cram, we replace the next hop arrays with a more compact data structure by either compressing with TCAM~({\em I1}) or compressing with SRAM~({\em I3}). Since the 25\% memory penalty of d-left hashing~\cite{dleft} is less expensive than TCAM's 3X higher area cost, we compress with SRAM by replacing the next hop arrays with a single SRAM-based hash table.

{\bf 3.} \sail uses a special next hop array $N_{32}$ for prefixes of length $>24$, which are very uncommon. Its entries are prefix expanded to 32 bits. In the worst case, a single prefix may be expanded into $2^{8}$ duplicate next hops. We address this memory inefficiency by replacing $N_{32}$ with a look-aside TCAM~({\em I6}) that can store prefixes of length $>24$ without additional expansion.

{\bf 4.} The number of bitmaps serves as a trade-off between the amount of parallelism required and the hash table's memory footprint. For \resail, we introduce a parameter $min\_bmp$ that represents the smallest bitmap available. In \cref{fig:resail}, $min\_bmp$ is 0 which means there are a total of 25 bitmaps from $B_{24}$ down to $B_0$. Increasing $min\_bmp$ reduces the number of parallel lookups at the cost of increased SRAM usage.

\subsection{Building the Data Structures} \label{subsec:building_resail}

\begin{table}[t]
    \centering
    \small
    \setlength\extrarowheight{1.1pt}
    \begin{tabular}{ccc}
    \hline
    \thead[c]{Entry} & \thead[c]{Prefix (Ternary)} & \thead[c]{Output Port} \\
        \hline
    1 & 010100** & A \\
    2 & 011***** & B \\
    3 & 100100** & C \\
    4 & 100101** & D \\
    5 & 10010100 & A \\
    6 & 10011010 & B \\
    7 & 10011011 & C \\
    8 & 10100011 & A \\
    \bottomrule
    \end{tabular}
    \caption{Example routing table}
    \label{tab:routing_table}
    \vspace{-\baselineskip}
\end{table}

\begin{table}[t]
    \centering
    \small
    \setlength\extrarowheight{1.1pt}
    \begin{tabular}{|c|c|c|}
    \hline
    \thead[c]{Index} & \thead[c]{Key} & \thead[c]{Value} \\
    \hline
    0 & 1001001 & C \\
    1 & 0101001 & A\\
    2 & 0111000 & B \\
    3 & - & - \\
    4 & 1001011 & D \\
    \bottomrule
    \end{tabular}
    \caption{Hash table for \cref{tab:routing_table} (pivot level = 6)}
    \label{tab:hash_table}
    \vspace{-\baselineskip}
\end{table}

{\bf Look-aside TCAM:}
Given a routing database, add all prefixes longer than 24 bits to the look-aside TCAM. Since there are very few IPv4 prefixes of length $> 24$, little TCAM is used.

{\bf Bitmaps:} \label{subsec:building_bitmaps}
For $i=24$ down to $i=min\_bmp$, construct a bitmap ($B_i$) of length $2^i$ such that every prefix of length $i$ in the routing database is marked as a 1 at the corresponding index. If $min\_bmp$ is not equal to 0, use prefix expansion~\cite{prefix_expansion} to combine $B_0$ to $B_{min\_bmp-1}$ into $B_{min\_bmp}$. Start with length $min\_bmp-1$ prefixes and work down linearly to length 0. A bit in $B_{min\_bmp}$ is flipped from 0 to 1 only if the bit is already a 0. This prevents incorrectly overwriting longer prefixes.

{\bf Hash Table:} \label{subsec:building_hash_table} We use d-left~\cite{dleft} for the hash table because it has a low probability of collision even when the ratio of entries to memory is as high as $80\%$. Hashing the matched prefix directly would require a separate hash table for each length from $min\_bmp$ to $24$, greatly fragmenting memory. Instead, \resail uses a standard trick~\cite{network_algorithmics} we call bit marking that enables us to generate hash keys of a fixed length. When an entry is added to bitmap $B_{i}$, its unique 25-bit hash key is produced by appending a $1$ and left shifting by $24-i$ bits. Each hash key is paired with its next hop and inserted into the hash table. The boundary of each prefix can be determined by scanning from the right for the first $1$. In effect, bit marking removes the need for multiple hash tables.

\cref{tab:hash_table} shows a hash table using 7-bit hash keys for \cref{tab:routing_table}. For simplicity, this example assumes a pivot level of 6 and a maximum prefix length of 8. Since entries 5-8 from \cref{tab:routing_table} are longer than the pivot length, they are not placed into the hash table (they are in the look-aside TCAM instead). The hash table has a size of 5 due to d-left's 25\% memory penalty. 011, a 3-bit entry, is appended with a 1 and left shifted 3 times, thus resulting in the hash key 0111000.

\subsection{Performing Lookups} \label{subsec:resail_lookups}

Start by performing two sets of lookups in parallel:
\begin{inlist}
  \item In the look-aside TCAM, perform a longest prefix match with the full 32-bit IPv4 address.
  \item From $B_{24}$ ($i=24$) down to $B_{min\_bmp}$, perform exact match lookups using the first $i$ bits of the destination address to directly index into $B_{i}$.
\end{inlist}

If a match is found in the look-aside TCAM, return its associated next hop. Otherwise, take the longest match across all bitmaps and generate its 25-bit hash key by bit marking.

At this step, either the next hop for a prefix match greater than 24 bits has been found or the final hash key has been created. In the latter case, use the hash key to perform an exact match lookup into the hash table to retrieve the next hop. 

\cref{alg:resail_lookup} in \cref{subsec:pseudocode} contains pseudocode for \resail lookups. \cref{subsubsec:resailupdates} describes incremental updates, deletions, and insertions in \resail.
\section{\bsic}
\label{sec:bsic}

\begin{figure}[t]
     \centering
     \begin{subfigure}[b]{0.49\textwidth}
         \centering
         \includegraphics[width=0.75\textwidth]{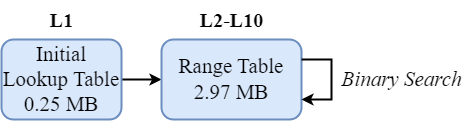}
         \caption{\cram model representation of \dxr ($k$=16)}
         \label{fig:dxr}
     \end{subfigure}
     \begin{subfigure}[b]{0.49\textwidth}
         \centering
         \includegraphics[width=0.85\textwidth]{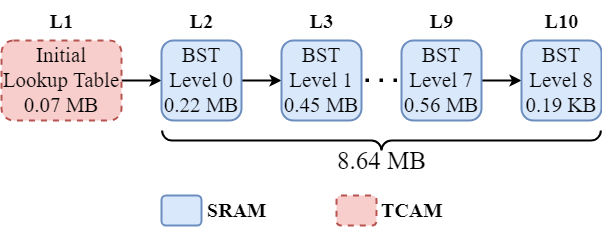}
         \caption{\cram model representation of \bsic ($k$=16)}
         \label{fig:bsic}
     \end{subfigure}
    \caption{\dxr vs \bsic for IPv4 prefixes in AS65000}
    \label{fig:dxrvsbsic}
    \vspace{-\baselineskip}
\end{figure}

By applying the optimization idioms to \dxr~\cite{dxr}, we create a new \cram lookup algorithm called \bsic, capable of supporting both IPv4 and IPv6. Refer to ~\cref{fig:dxrtobsic} for the intuition to which we now add details.

{\bf \dxr Review:} Inspired by \cite{dir} and \cite{lampson}, \dxr performs binary search on range endpoints using a range table~(binary search table). To reduce the depth of binary search, \dxr uses an initial lookup table directly indexed by the first $k$ bits of the address. The lookup table returns a pointer to the subsection of the range table in which binary search will be performed. \dxr adds two optimizations:
\begin{inlist}
  \item Merging neighboring ranges that point to the same next hop.
  \item Discarding right endpoints.
\end{inlist}

\subsection{Applying the Optimization Idioms} \label{subsec:designing_bsic}

We show the \cram derivation of \bsic using the IPv4 AS65000 BGP routing table (Sep 2023). Start with the \cram model representation of \dxr in \cref{fig:dxr}. As recommended by \cite{dxr}, set $k=16$~(D16R) for the best IPv4 results. To obtain \bsic in \cref{fig:bsic}, use the idioms as follows:

{\bf 1.} Since \dxr's initial lookup table relies on direct indexing, leaving many entries unused, we compress with TCAM ({\em I1}). Replacing the SRAM-based initial lookup table with a TCAM-based one reduces its memory consumption by over 3X, from 0.25 MB of SRAM to 0.07 MB of TCAM.

{\bf 2.} In \dxr, the range table is repeatedly accessed during binary search. Since lookup tables are limited to a single access per packet in the \cram model, the range table must be split up. We do so through memory fan-out ({\em I8}). By converting the range table into multiple binary search trees (BSTs) and distributing search levels across separate tables accessed at different steps, we ensure each table is visited at most once per packet. However, this greatly increases the amount of memory needed because every internal node has to store up to two pointers. In \cref{fig:dxrvsbsic}, \dxr's range table uses only 2.97 MB of SRAM while \bsic's BST levels use 8.64 MB of SRAM---a 2.9X increase. Although costly, memory fan-out is essential because duplicating the entire range table for each search level would require an infeasible amount of SRAM (26.73 MB).

{\bf 3.} The parameter $k$ is a strategic cut ({\em I4}) that balances memory usage in the initial lookup table against the number of required BST levels. Due to the high memory cost of direct indexing, \dxr's SRAM-based lookup table is limited to $k<=20$. For example, if $k=24$, \dxr's initial lookup table would consume 64 MB of SRAM. To effectively support IPv6, which has longer prefixes, a larger $k$ value is required. Since TCAM can store wildcard entries without prefix expansion, \bsic's TCAM-based initial lookup table can use much larger $k$ values, up to the underlying TCAM block width ($k=44$ for Tofino-2).

\subsection{Building the Data Structures} \label{subsec:building_bsic}

\begin{table}[t]
    \centering
    \small
    \setlength\extrarowheight{1.1pt}
    \begin{tabular}{|c|c|c|}
    \hline
    \thead[c]{Key} & \thead[c]{Value} & \thead[c]{BST Entries (for reference)} \\
    \hline
    0101 & Pointer to BST 1 & 00** \\
    011* & Next Hop B & -\\
    1001 & Pointer to BST 2 & 00**, 01**, 0100, 1010, 1011 \\
    1010 & Pointer to BST 3 & 0011 \\
    \bottomrule
    \end{tabular}
    \caption{Initial lookup table for \cref{tab:routing_table} ($k=4$)}
    \label{tab:initial_lookup_table}
    \vspace{-\baselineskip}
\end{table}

{\bf Initial Lookup Table:}
Given a routing database and a slice size $k$, populate the initial lookup table by storing all prefixes as unique $k$-length slices. Duplicate slices are condensed into one entry. Three cases arise when adding a prefix $p$ of length $l$:

\begin{enumerate}
  \item If $l$ < $k$, pad $p$ with $k-l$ wildcard (*) bits. Its associated table value is $p$'s next hop.
  \item If $l$ \texttt{==} $k$, do not modify $p$. If there are longer prefixes that share the same $k$-length slice as $p$, its associated table value is a pointer to the corresponding BST's root node. Otherwise, its associated table value is $p$'s next hop.
  \item If $l$ > $k$, trim $p$ down to $k$ bits. Its associated table value is a pointer to the corresponding BST's root node.
\end{enumerate}

\cref{tab:initial_lookup_table} shows an initial lookup table with $k$=4 created using \cref{tab:routing_table}. The maximum prefix length is 8. The Key column contains all the $k$-length slices while the Value column stores the associated pointers and next hops. The BST Entries column shows the prefix segments that are pointed to by the slices. Since entries 3-7 in \cref{tab:routing_table} share the same $k$-length slice, they are condensed into a single key 1001 that points to BST 2.

{\bf Binary Search Trees (BSTs):}
To create a BST for a given lookup table entry, identify all prefixes in the database that match the entry up to the $k$th bit. For all such prefixes, store the remaining bits and next hops as tuples in a list. Take the list of tuples and perform the range expansion and optimizations described in \cite{dxr}. We defer these details to \cref{subsec:bsic_details}.

Use the resulting list of left endpoints to construct a BST in which every node contains four fields: pointers to the left and right child, the next hop, and the left endpoint itself. Repeat this process for all lookup table entries containing pointers.

\subsection{Performing Lookups} \label{subsec:bsic_lookups}
Start in the initial lookup table by performing a longest prefix match using the first $k$ bits of the destination address. If either a next hop is returned or a miss occurs, search terminates. If a pointer to a BST is returned, follow the pointer to the corresponding root node and form the next search key by extracting the remaining bits of the destination address. Once at a node, perform standard binary search using the search key.

\cref{alg:bsic_lookup} in \cref{subsec:pseudocode} contains pseudocode for \bsic lookups. \cref{subsubsec:bsicupdates} describes incremental updates, deletions, and insertions in \bsic.
\section{\mashup}
\label{sec:mashup}

\begin{figure}[t]
     \centering
     \begin{subfigure}[b]{0.49\textwidth}
         \centering
         \includegraphics[width=0.58\textwidth]{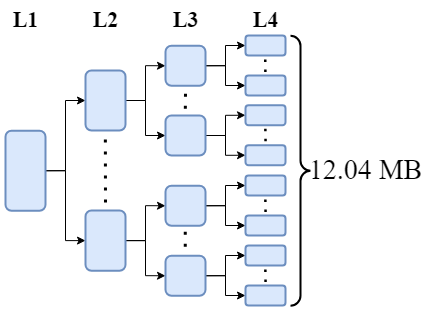}
         \caption{\cram model representation of multibit trie (16-4-4-8)}
         \label{fig:multibit}
     \end{subfigure}
     \begin{subfigure}[b]{0.49\textwidth}
         \centering
         \includegraphics[width=0.68\textwidth]{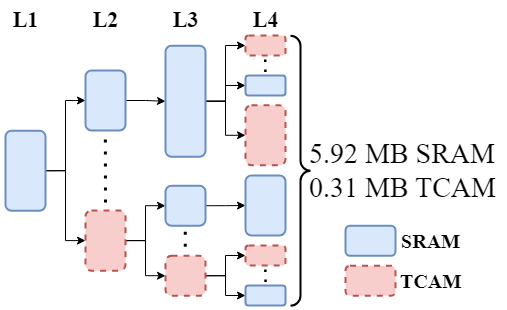}
         \caption{\cram model representation of \mashup (16-4-4-8)}
         \label{fig:mashup}
     \end{subfigure}
    \caption{Multibit trie vs \mashup for IPv4 prefixes in AS65000}
    \label{fig:multibitvsmashup}
    \vspace{-\baselineskip}
\end{figure}

By applying the optimization idioms to multibit tries~\cite{prefix_expansion}, we create a new \cram lookup algorithm called \mashup, capable of supporting both IPv4 and IPv6. Refer to ~\cref{fig:multibittomashup} for the intuition to which we now add details.

{\bf Multibit Trie Review:} Multibit tries are search tries that examine multiple bits per lookup, known as a stride. Reducing the number of strides decreases the number of worst-case memory accesses, but increases prefix expansion and memory usage. We assume each tree level has exactly one stride.

\subsection{Applying the Optimization Idioms} \label{subsec:designing_mashup}

We show the \cram derivation of \mashup using the IPv4 AS65000 BGP routing table (Sep 2023). Start with the \cram model representation of a multibit trie in \cref{fig:multibit}. We find strides 16-4-4-8 yield the best IPv4 results (\cref{subsec:parameters}). To obtain \mashup in \cref{fig:mashup}, use the idioms as follows:

{\bf 1.} For each trie node, we consider both compressing with TCAM ({\em I1}) and expanding to SRAM ({\em I2}). If the increase in memory due to prefix expansion~\cite{prefix_expansion} is less than 3X, we use SRAM. Otherwise, we use TCAM. This results in a hybrid trie with both TCAM and SRAM nodes, as seen in \cref{fig:mashup}.

{\bf 2.} Once the trie is hybridized, apply table coalescing ({\em I5}) by merging partially filled nodes\footnote{For \mashup in \cref{subsec:before_implementation}, we greedily fill the largest tables with the smallest ones. This approach is easy to implement, but possibly suboptimal.} of the same memory type into super-tables, compactly mapping them onto contiguous TCAM blocks or SRAM pages with minimal fragmentation. This requires prepending entries with a tag~\cite{tag_bits} to distinguish between logical tables. A tag of width $x$ can coalesce $2^x$ logical tables into one super-table. The combination of node hybridization and table coalescing reduces SRAM usage from 12.04 MB to 5.92 MB at the cost of 0.31 MB of TCAM (\cref{fig:multibitvsmashup}).

{\bf 3.} The set of strides is a parameter that serves as a strategic cut ({\em I4}). For a given set of strides, the trie's memory overhead is directly proportional to the number of internal pointers. A simple method for choosing strides, explained in \cref{subsec:parameters}, is to analyze the database's prefix length distribution.

We omit standard algorithms for building the \mashup trie, as the process is identical to constructing a multibit trie, which has been extensively studied in prior work~\cite{prefix_expansion, network_algorithmics}.

\subsection{Performing Lookups} \label{subsec:mashup_lookups}

Let $S_i$ represent the $i$-th stride. Start in the root node by performing a lookup with the first $S_0$ bits of the destination address. If the current node is TCAM, perform a longest prefix match. Otherwise, perform an exact match. If a miss occurs, terminate the search. If a hit occurs, three values may be returned: a next hop, a pointer to the next node, and a unique tag. If a next hop is returned, save it. Form the lookup key for level $i$ by extracting the next $S_i$ bits of the destination address and prepending the current tag. Repeat the lookup process until either a leaf node is reached or a miss occurs. Upon termination, return the saved next hop.

\cref{alg:mashup_lookup} in \cref{subsec:pseudocode} contains pseudocode for \mashup lookups. \cref{subsubsec:mashupupdates} describes incremental updates, deletions, and insertions in \mashup.
\section{Results}
\label{sec:results}

\begin{figure}[t]
\centering
\begin{tikzpicture}
    \large
    \begin{axis}[
        xlabel=Prefix Length,
        ylabel=\% of Database,
        xmin=0, xmax=64,
        ymin=0, ymax=70,
        xtick={0,4,...,64},
        ytick={0,10,...,70},
        width=\columnwidth,
        height=110pt,
        xticklabel style={rotate=0,font=\small},
        yticklabel style={rotate=0,font=\small},
        xlabel style={font=\normalsize,at={(axis description cs:.5,-.2)}},
        ylabel style={font=\normalsize,at={(axis description cs:-.075,.5)}},
        legend style={at={(1,1)},anchor=north east,font=\small},
        tick pos=left
    ]
    \addplot[line width=1.25pt,mark=none,smooth,color=blue]
    plot coordinates {
        (0,0)
        (10,0)
        (11,0.01)
        (12,0.03)
        (13,0.06)
        (14,0.13)
        (15,0.23)
        (16,1.45)
        (17,0.89)
        (18,1.49)
        (19,2.7)
        (20,4.78)
        (21,5.54)
        (22,11.84)
        (23,10.68)
        (24,60.07)
        (25,0.08)
        (26,0)
        (32,0)
    };
    \addlegendentry{IPv4}
    \addplot[line width=1.25pt,mark=none,densely dashed,color=red]
    plot coordinates {
        (0,0)
        (19,0)
        (20,0.01)
        (21,0)
        (22,0)
        (23,0)
        (24,0.02)
        (25,0)
        (26,0.01)
        (27,0.01)
        (28,0.11)
        (29,2.31)
        (30,0.32)
        (31,0.16)
        (32,12.17)
        (33,2.07)
        (34,1.96)
        (35,0.69)
        (36,3.53)
        (37,0.57)
        (38,0.94)
        (39,0.77)
        (40,7.91)
        (41,0.51)
        (42,1.9)
        (43,0.65)
        (44,9.27)
        (45,1.16)
        (46,2.54)
        (47,2.97)
        (48,46.48)
        (49,0.01)
        (50,0)
        (51,0)
        (52,0.03)
        (53,0)
        (55,0)
        (56,0.44)
        (57,0)
        (63,0)
        (64,0.46)
    };
    \addlegendentry{IPv6}
    \end{axis}
\end{tikzpicture}
\vspace{-0.5em}
\caption{IPv4 and IPv6 prefix length distributions in AS65000 and AS131072, respectively, for September 2023}
\label{fig:bgp_dist}
\vspace{-\baselineskip}
\end{figure}
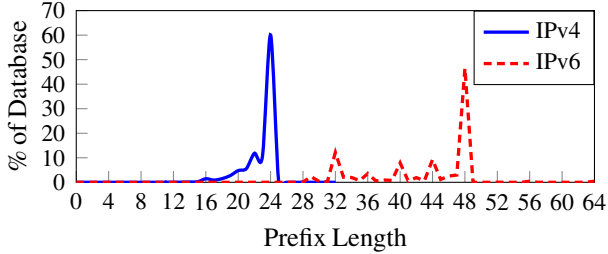

In this section, we introduce the databases, target implementations, and parameter values that we selected for our experiments. We use the \cram metrics to determine the best IPv4 (\resail) and IPv6 (\bsic) algorithms {\em before} implementation. We then compare the resource utilizations of the best \cram algorithms with those of state-of-the-art IP lookup schemes {\em after} implementation.

\subsection{Databases} \label{subsec:databases}

For IPv4, we used the AS65000 BGP routing table (Sep. 2023)~\cite{cidr_report_ipv4} which has close to 930k IPv4 prefixes. For IPv6, we used the AS131072 BGP routing table (Sep. 2023)~\cite{cidr_report_ipv6} which has close to 190k IPv6 prefixes. \cref{fig:bgp_dist} shows their prefix length distributions. We identify three key patterns:

{\bf P1.} Prefix distributions have major and minor spikes. For IPv4, there is a major spike at length 24 and minor spikes at lengths 16, 20, and 22. For IPv6, there is a major spike at length 48 and minor spikes at lengths 28, 32, 36, 40, and 44.

{\bf P2.} The majority of IPv4 prefixes are longer than 12 bits.

{\bf P3.} The majority of IPv6 prefixes are longer than 28 bits.

\subsection{Target Implementations} \label{subsec:target_implementations}

We obtained results for two different targets: an ideal RMT chip and Intel Tofino-2 (also RMT). Since RMT is a stricter variant of dRMT with additional access restrictions, we expect our RMT results to be reproducible on a dRMT chip.

{\bf Ideal RMT Chip (Simulation):} We define an ideal RMT chip to be an RMT chip with Tofino-2 specifications (same memory, number of stages, etc.)~\cite{bf_p4c} that can achieve 100\% SRAM utilization and perform at least two dependent ALU operations per stage. The resource utilization for an ideal RMT chip is obtained through simulation by using Tofino-2 SRAM page~(128x1024b) and TCAM block~(44x512b) sizes. If the number of TCAM blocks or SRAM pages used by a table exceeds the amount available in a MAU~(stage), the table is simply partitioned across multiple MAUs. Since Tofino-2 has 20 MAUs, results that require over 20 are considered infeasible.


{\bf Intel Tofino-2 (Implementation):} The resource utilization for Tofino-2 is obtained through implementation. We implement the best \cram algorithms using P4 and compile them with the Intel P4 compiler. P4 Insight~\cite{p4_insight} then outputs detailed resource mappings and visualizations specific to Tofino-2.

\subsection{Parameter Values} \label{subsec:parameters}

We choose parameter values based on observations from \cref{subsec:databases}.

\resail's key parameter is $min\_bmp$, the smallest bitmap available. We choose $min\_bmp=13$ because there are so few IPv4 prefixes shorter than 13 bits {\bf (P2)}, thus minimizing the amount of prefix expansion needed.

\bsic's key parameter is $k$, the initial slice size. As recommended by \cite{dxr}, for IPv4, we choose $k=16$. For IPv6, we choose $k=24$ because most IPv6 prefixes are longer than 28 bits {\bf (P3)}. Therefore, a $k$ value that is close to but smaller than 28 can compress over 190k prefixes into just 7k TCAM entries. We briefly explore other choices of $k$ for IPv6 and examine potential latency-memory trade-offs in \cref{subsec:tradeoff}.

\mashup's key parameter is its set of strides. Intuitively, we want to select strides that mirror the distribution spikes {\bf (P1)} seen in \cref{fig:bgp_dist} because they will minimize prefix expansion. For IPv4, we choose 16-4-4-8~(spikes at 16, 20, 24). For IPv6, we choose 20-12-16-16~(spikes at 32, 48). We do not select 32 as the first stride because it is too wide---especially for the root node which may contain many entries. Therefore, we decompose 32 into separate strides of 20 and 12.

\subsection{Comparisons {\em before} Implementation} \label{subsec:before_implementation}

\begin{table}[t]
    \centering
    \small
    \setlength\extrarowheight{1.25pt}
    \setlength\tabcolsep{2.5pt}
    \begin{tabular}{lccc}
    \hline
      \thead[l]{Scheme} & \thead[c]{TCAM Bits} & \thead[c]{SRAM Bits} & \thead[c]{Steps} \\ 
        \hline
        \mashup (16-4-4-8) & 0.31 MB & 5.92 MB & 4 \\
        \bsic ($k$=16) & 0.07 MB & 8.64 MB & 10 \\
        \resail ($min\_bmp$=13) & 3.13 KB & 8.58 MB & 2 \\
        \bottomrule
    \end{tabular}
    \vspace{-0.5em}
    \caption{\cram metrics for IPv4 prefixes in AS65000}
    \label{tab:cram_ipv4}
    \vspace{-0.5em}
\end{table}

\begin{table}[t]
    \centering
    \small
    \setlength\extrarowheight{1.25pt}
    \setlength\tabcolsep{2.5pt}
    \begin{tabular}{lccc}
    \hline
      \thead[l]{Scheme} & \thead[c]{TCAM Bits} & \thead[c]{SRAM Bits} & \thead[c]{Steps} \\ 
        \hline
        \mashup (20-12-16-16) & 0.32 MB & 0.77 MB & 4 \\
        \bsic ($k$=24) & 0.02 MB & 3.18 MB & 14 \\
        \bottomrule
    \end{tabular}
    \vspace{-0.5em}
    \caption{\cram metrics for IPv6 prefixes in AS131072}
    \label{tab:cram_ipv6}
    \vspace{-\baselineskip}
\end{table}

Recall that \cram metrics enable quick estimation of algorithm scalability {\em before} implementation.

We present IPv4 and IPv6 \cram metrics in \cref{tab:cram_ipv4} and \cref{tab:cram_ipv6}, derived from the optimization steps in \cref{subsec:designing_resail}, \cref{subsec:designing_bsic}, and \cref{subsec:designing_mashup}, for the IPv4 and IPv6 BGP tables, respectively.

For IPv4, \resail outperforms \bsic in all three \cram metrics. Between \resail and \mashup, \resail wins in TCAM and steps but loses in SRAM. However, \mashup requires 100X more TCAM than \resail, whereas \resail requires only 1.4X more SRAM than \mashup. Therefore, we determine \resail to be the best \cram IPv4 algorithm.

For IPv6, we choose between \bsic and \mashup. \bsic wins in TCAM but loses in SRAM and steps. \mashup requires 16X more TCAM than \bsic, while \bsic requires roughly 4X more SRAM and steps than \mashup. As before, we prioritize TCAM because it is more expensive and limited than SRAM---for example, Tofino-2 contains 19X more SRAM than TCAM. Although \bsic uses more steps than \mashup, this is due to \bsic's use of BSTs, which have a high initial step cost. Therefore, we determine \bsic to be the best \cram IPv6 algorithm {\em for Tofino-2}. However, for more stage-constrained ASICs, \mashup is likely better.

\begin{table}[t]
    \centering
    \small
    \setlength\extrarowheight{1.25pt}
    \setlength\tabcolsep{2.5pt}
    \begin{tabular}{lccc}
    \hline
      \thead[l]{Scheme} & \thead[c]{TCAM Blocks} & \thead[c]{SRAM Pages} & \thead[c]{Stages} \\ 
        \hline
        \mashup (16-4-4-8) & 235 & 216 & 10 \\
        \bsic ($k$=16) & 74 & 558 & 16 \\
        \resail ($min\_bmp$=13) & 2 & 556 & 9 \\
        \bottomrule
    \end{tabular}
    \vspace{-0.5em}
    \caption{Ideal RMT mapping for IPv4 prefixes in AS65000}
    \label{tab:ideal_rmt_ipv4}
    \vspace{-0.5em}
\end{table}

\begin{table}[t]
    \centering
    \small
    \setlength\extrarowheight{1.25pt}
    \setlength\tabcolsep{2.5pt}
    \begin{tabular}{lccc}
    \hline
      \thead[l]{Scheme} & \thead[c]{TCAM Blocks} & \thead[c]{SRAM Pages} & \thead[c]{Stages} \\ 
        \hline
        \mashup (20-12-16-16) & 178 & 47 & 8 \\
        \bsic ($k$=24) & 15 & 211 & 14 \\
        \bottomrule
    \end{tabular}
    \vspace{-0.5em}
    \caption{Ideal RMT mapping for IPv6 prefixes in AS131072}
    \label{tab:ideal_rmt_ipv6}
    \vspace{-\baselineskip}
\end{table}

To verify the validity of the \cram metrics, we explicitly map each \cram algorithm to an ideal RMT chip and present its resource utilization in \cref{tab:ideal_rmt_ipv4} and \cref{tab:ideal_rmt_ipv6}. This mapping, which accounts for Tofino-2 TCAM block sizes, SRAM page sizes, and per-stage memory, is precisely the complicated process that the \cram model seeks to relieve algorithm designers of. Comparing \cref{tab:cram_ipv4} with \cref{tab:ideal_rmt_ipv4} and \cref{tab:cram_ipv6} with \cref{tab:ideal_rmt_ipv6}, observe that the \cram metrics accurately predict a target algorithm's resource utilization and potential scalability.

\subsection{Comparisons {\em after} Implementation}
\label{subsec:after_implementation}

The previous subsection compared our three new algorithms {\em before} implementation using \cram metrics. Here, we compare the best \cram algorithms {\em after} implementation on Tofino-2 against the best pre-existing IPv4 and IPv6 schemes.

\subsubsection{Baseline Selection} \label{subsubsec:baselines}

We select four single-resource baselines: SRAM-only for IPv4 and IPv6, and TCAM-only for IPv4 and IPv6.

{\bf SRAM-only for IPv4:} We choose \sail~\cite{sail} as our SRAM-only IPv4 baseline due to its on-chip memory bound for short prefixes, which enables it to scale very well. Although IPv4 schemes like Poptrie~\cite{poptrie} and \dxr~\cite{dxr} use less memory, they require too many memory accesses and stages.

{\bf SRAM-only for IPv6:} We choose \hibst~\cite{hibst} as our SRAM-only IPv6 baseline because it is the most memory-efficient IPv6 lookup algorithm to date~\cite{xoroffsettrie}. It uses a treap data structure that maps each prefix to a unique node.

{\bf TCAM-only for IPv4 and IPv6:} We choose a logical TCAM as our TCAM-only IPv4 and IPv6 baseline because, although TCAM-oriented schemes exist for reducing power consumption~\cite{liu} or merging multiple FIBs~\cite{virtual_routers}, none focus on scaling IP lookup for a single database.

\begin{table}[t]
    \centering
    \small
    \setlength\extrarowheight{1.25pt}
    \setlength\tabcolsep{2.5pt}
    \begin{tabular}{lcccc}
    \hline
      \thead[l]{Scheme} & \thead[c]{TCAM \\ Blocks} & \thead[c]{SRAM \\ Pages} & \thead[c]{Stages} & \thead[c]{Target Chip} \\ 
        \hline
        \resail ($min\_bmp$=13) & 17 & 750 & 16 & Tofino-2 \\
        \resail ($min\_bmp$=13) & 2 & 556 & 9 & Ideal RMT \\
        \sail & - & 2313 & 33 & Ideal RMT \\
        Logical TCAM & 1822 & - & 76 & Ideal RMT \\
        Tofino-2 Pipe Limit & 480 & 1600 & 20 & - \\
        \bottomrule
    \end{tabular}
    \vspace{-0.5em}
    \caption{Baseline comparison for IPv4 prefixes in AS65000}
    \label{tab:after_implementation_ipv4}
    \vspace{-0.5em}
\end{table}

\begin{table}[t]
    \centering
    \small
    \setlength\extrarowheight{1.25pt}
    \setlength\tabcolsep{2.5pt}
    \begin{tabular}{lcccc}
    \hline
      \thead[l]{Scheme} & \thead[c]{TCAM \\ Blocks} & \thead[c]{SRAM \\ Pages} & \thead[c]{Stages} & \thead[c]{Target Chip} \\ 
        \hline
        \bsic ($k$=24) & 15 & 416 & 30 & Tofino-2 \\
        \bsic ($k$=24) & 15 & 211 & 14 & Ideal RMT \\
        \hibst & - & 219 & 18 & Ideal RMT \\
        Logical TCAM & 762 & - & 32 & Ideal RMT \\
        Tofino-2 Pipe Limit & 480 & 1600 & 20 & - \\
        \bottomrule
    \end{tabular}
    \vspace{-0.5em}
    \caption{Baseline comparison for IPv6 prefixes in AS131072}
    \label{tab:after_implementation_ipv6}
    \vspace{-\baselineskip}
\end{table}

\subsubsection{IPv4 Comparison} \label{subsubsec:ipv4_comparison}

\cref{tab:after_implementation_ipv4} compares the ideal RMT resource utilization of \resail with the IPv4 baselines. \resail requires 911X fewer TCAM blocks than the logical TCAM and approximately 4X fewer SRAM pages and stages than \sail. Although \sail's memory scales efficiently, its high upfront cost makes it impractical for RMT-like chips. \resail outperforms the logical TCAM, which only supports IPv4 databases of up to 245,760 entries---about 3.8X smaller than the current IPv4 BGP table.

\cref{tab:after_implementation_ipv4} also compares \resail on an ideal RMT chip and on Tofino-2. While \resail fits on Tofino-2 for AS65000, it requires nearly 1.4X more SRAM pages and 2X more stages than on an ideal RMT chip. This is because Tofino-2 reserves bits in each SRAM word for identifying actions, limiting the maximum SRAM utilization to 50\%. The increase in TCAM is due to extra ternary bitmask tables needed for extracting bits.

\subsubsection{IPv6 Comparison} \label{subsubsec:ipv6_comparison}

\cref{tab:after_implementation_ipv6} compares the ideal RMT resource utilization of \bsic with the IPv6 baselines. \bsic uses less SRAM and fewer stages than \hibst, at the cost of 15 TCAM blocks. Both \bsic and \hibst support the current IPv6 BGP table, whereas the logical TCAM only supports up to 122,880 entries---about 1.6X smaller than the current IPv6 BGP table.

\cref{tab:after_implementation_ipv6} also compares \bsic on an ideal RMT chip and on Tofino-2. Our ideal RMT chip assumes each stage can perform at least two dependent ALU operations. However, in practice, a Tofino-2 stage can execute only one level of ALU logic. Consequently, each BST level requires two stages: one for comparing the search key and another for performing the P4 action. This creates a repeating pattern where an SRAM-intensive stage is followed by a stage with minimal SRAM usage. Although \bsic on Tofino-2 requires 30 stages (ten over the Tofino-2 pipe limit), we successfully fit \bsic for AS131072 by recirculating each packet. However, this effectively halves the number of available switch ports.
\section{Scalability}
\label{sec:scalability}

While \cref{sec:results} presents results for {\em current} BGP tables, this section presents scalability analysis for \resail and \bsic on larger {\em synthetic} routing databases. We omit scalability analysis for \mashup because it requires too much TCAM (for Tofino-2).

\subsection{IPv4 Scaling} \label{subsec:ipv4_scaling}

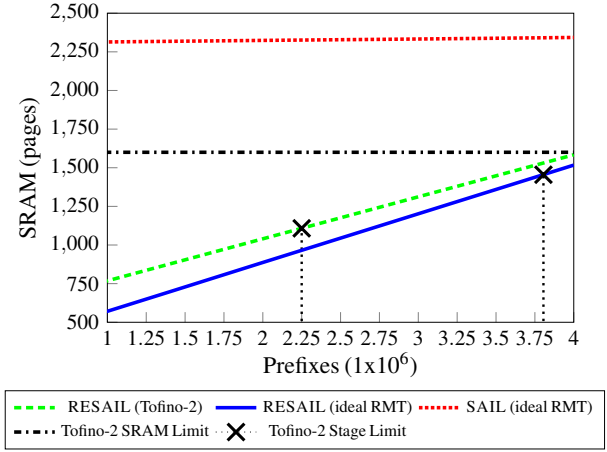
\begin{figure}
\centering
\begin{tikzpicture}[scale=0.9]
    \normalsize
    \begin{axis}[
        xlabel=Prefixes (1x$10^6$),
        ylabel=SRAM (pages),
        xmin=1, xmax=4,
        ymin=500, ymax=2500,
        xtick={1,1.25,...,4},
        ytick={500,750,...,2500},
        width=\columnwidth,
        height=175pt,
        xticklabel style={rotate=0,font=\small},
        yticklabel style={rotate=0,font=\small},
        xlabel style={font=\normalsize,at={(axis description cs:.5,-.07)}},
        ylabel style={font=\normalsize,at={(axis description cs:-.125,.5)}},
        legend columns=3,
        legend style={at={(-0.22,-0.22)},anchor=north west,font=\scriptsize},
        tick pos=left
    ]
    
    \draw[dotted, black, line width=1pt] plot[domain=0:4] (3.805000, 0) -- (3.805000, 1455);
    \draw[dotted, black, line width=1pt] plot[domain=0:4] (2.250000, 0) -- (2.250000, 1108);

    \addplot[line width=1.5pt,mark=none,densely dashed,color=green,domain=0:4] {271.8*x + 496.5};
    \addlegendentry{\resail (Tofino-2)}
    
    \addplot[line width=1.5pt,mark=none,smooth,color=blue]
    plot coordinates {
        (0.932861,556)
        (1.000000,571)
        (2.000000,887)
        (3.000000,1202)
        (3.800000,1454)
        (4.000000,1517)
    };
    \addlegendentry{\resail (ideal RMT)}
    \addplot[line width=1.5pt,mark=none,densely dotted,color=red]
    plot coordinates {
        (0.932861,2313)
        (1.000000,2314)
        (2.000000,2324)
        (3.000000,2333)
        (3.800000,2341)
        (4.000000,2343)
    };
    \addlegendentry{\sail (ideal RMT)}

    \addplot[line width=1.5pt,mark=none,dash pattern=on 3pt off 2pt on 1pt off 2pt,color=black,domain=0:4] {1600};
    \addlegendentry{Tofino-2 SRAM Limit}

    \addplot[dotted,mark=x,mark size=5pt,mark options={solid, line width=1.3pt}] coordinates {(2.250000, 1108)};
    \addplot[dotted,mark=x,mark size=5pt,mark options={solid, line width=1.3pt}] coordinates {(3.805000, 1455)}; 
    \addlegendentry{Tofino-2 Stage Limit}
    
    \end{axis}
\end{tikzpicture}
\caption{\resail vs \sail scaling (IPv4)}
\label{fig:ipv4_scaling}
\vspace{-\baselineskip}
\end{figure}

\cref{fig:ipv4_scaling} shows IPv4 scalability results for \resail and our SRAM-only IPv4 baseline, \sail. We did not generate synthetic prefixes because the resource utilization of \resail and \sail depends on the distribution of {\em prefix lengths} rather than the distribution of the prefixes themselves. From the perspective of memory usage, \resail and \sail do not distinguish between prefixes of identical length. Therefore, we use a simple scaling model that applies a constant scaling factor to all prefix lengths. For the ideal RMT results of \resail and \sail, we use the steps described in \cref{subsec:target_implementations} to calculate their new utilization. For the Tofino-2 results of \resail, we update the corresponding P4 table sizes to reflect the larger databases.

For ideal RMT, \sail is infeasible because its SRAM cost far exceeds the Tofino-2 SRAM limit. At any given database size, \resail for Tofino-2 uses more SRAM than \resail for ideal RMT. This is expected since Tofino-2 does not allow 100\% SRAM utilization. Notably, \resail on an ideal RMT chip scales to around 3.8 million prefixes, 4X larger than the current IPv4 BGP table. \resail on Tofino-2 scales to around 2.25 million prefixes, 2.3X larger than the current routing database and significantly beyond \sail's capacity.

\subsection{IPv6 Scaling} \label{subsec:ipv6_scaling}

\begin{figure}
\centering
\begin{tikzpicture}[scale=0.9]
    \normalsize
    \begin{axis}[
        xlabel=Prefixes (1x$10^5$),
        ylabel=SRAM (pages),
        xmin=2, xmax=7,
        ymin=200, ymax=1700,
        xtick={2,2.5,...,7},
        ytick={200,350,...,1700},
        width=\columnwidth,
        height=175pt,
        xticklabel style={rotate=0,font=\small},
        yticklabel style={rotate=0,font=\small},
        xlabel style={font=\normalsize,at={(axis description cs:.5,-.07)}},
        ylabel style={font=\normalsize,at={(axis description cs:-.125,.5)}},
        legend columns=3,
        legend style={at={(-0.22,-0.22)},anchor=north west,font=\scriptsize},
        tick pos=left
    ]
    
    \draw[dotted, black, line width=1pt] plot[domain=0:7] (6.32717, 0) -- (6.32717, 750);
    \draw[dotted, black, line width=1pt] plot[domain=0:7] (3.89269, 0) -- (3.89269, 810);
    \draw[dotted, black, line width=1pt] plot[domain=0:7] (3.40943, 0) -- (3.40943, 383);

    \addplot[line width=1.5pt,mark=none,densely dashed,color=green,domain=0:7] {208.9*x - 3.335};
    \addlegendentry{\bsic (Tofino-2)}
    
    \addplot[line width=1.5pt,mark=none,smooth,color=blue]
    plot coordinates {
        (1.94635,211)
        (3.89269,423)
        (5.83903,686)
        (6.16385,726)
        (6.32717,740)
        (6.81401,776)
        (7.78537,867)
    };
    \addlegendentry{\bsic (ideal RMT)}
    \addplot[line width=1.5pt,mark=none,densely dotted,color=red]
    plot coordinates {
        (1.94635,219)
        (3.89270,437)
        (5.83905,655)
        (6.36743,715)
        (7.13538,801)
    };
    \addlegendentry{\hibst (ideal RMT)}

    \addplot[line width=1.5pt,mark=none,dash pattern=on 3pt off 2pt on 1pt off 2pt,color=black,domain=0:7] {1600};
    \addlegendentry{Tofino-2 SRAM Limit}

    \addplot[dotted,mark=x,mark size=5pt,mark options={solid, line width=1.3pt}] coordinates {(6.32717, 750)};
    \addplot[dotted,mark=x,mark size=5pt,mark options={solid, line width=1.3pt}] coordinates {(3.89269, 810)};
    \addplot[dotted,mark=x,mark size=5pt,mark options={solid, line width=1.3pt}] coordinates {(3.40943, 383)};
    \addlegendentry{Tofino-2 Stage Limit}
    
    \end{axis}
\end{tikzpicture}
\caption{\bsic vs \hibst scaling (IPv6)}
\label{fig:ipv6_scaling}
\vspace{-\baselineskip}
\end{figure}
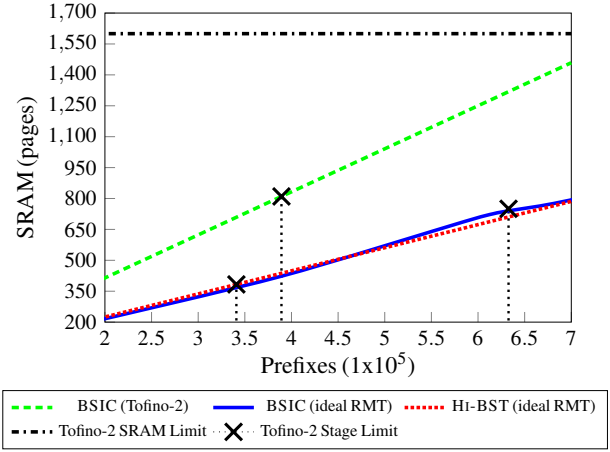

\cref{fig:ipv6_scaling} shows IPv6 scalability results for \bsic and our SRAM-only IPv6 baseline, \hibst. For \bsic, we generated synthetic prefixes because its resource utilization depends on the distribution of prefixes and sub-prefixes. To obtain worst-case scalability results, observe that the first three bits of IPv6 prefixes in AS131072 are 000---forming an {\em IPv6 universe}. We use different combinations of these bits to generate significantly larger synthetic databases from AS131072, an approach we call {\em multiverse scaling}. Multiverse scaling assumes that the distribution of all prefix lengths scales uniformly. In practice, customer scaling causes some prefixes (e.g., /48s) to scale more rapidly than others (e.g., /24s). However, this stresses only the BSTs and not the initial TCAM, unlike multiverse scaling which models worst-case results for TCAM, SRAM, and stages. For \hibst, we use the memory calculation provided in \cite{hibst}.

As seen in \cref{fig:ipv6_scaling}, both instances of \bsic are able to out-scale \hibst. For an ideal RMT chip, \hibst only scales to around 340k prefixes, 1.8X larger than the current IPv6 BGP Table. Even though \hibst is the most memory efficient IPv6 lookup scheme, it requires too many stages. Comparing the two instances of \bsic, we see that \bsic for ideal RMT scales to around 630k prefixes, 3.3X larger than the current routing database. Since \bsic for Tofino-2 uses over 2X more stages, it scales to around 390k prefixes---2X the size of the current IPv6 BGP table.
\section{\cram Model Evaluation}
\label{sec:crammodelevaluation}

How predictive was the \cram model? \cref{subsec:before_implementation} showed that the \cram model accurately predicted \resail and \bsic as the best algorithms for IPv4 and IPv6, respectively. We now examine the \cram metrics in more detail.


\cref{tab:ipv4_cram_evaluation} and \cref{tab:ipv6_cram_evaluation} show results for \resail and \bsic on three models: the \cram model, an ideal RMT model, and a Tofino-2 implementation. The three models form a hierarchy of abstractions with increasing detail. We scale the \cram metrics found in \cref{tab:cram_ipv4} and \cref{tab:cram_ipv6} from raw bits to TCAM blocks and SRAM pages to allow for uniform comparisons. The three models can be understood as follows:

\textbf{\cram model:} Using the \cram metrics (raw bits and dependent steps), an algorithm designer can quickly predict scalability {\em without} seeing the product data sheet.

\textbf{Ideal RMT model:} This model allows for more precise predictions but requires a {\em basic} understanding of the data sheet, specifically the general organization of memory and stages.

\textbf{Tofino-2 implementation:} This is the most accurate model but also the most complex to develop. It accounts for low-level details that are hard to glean from data sheets, such as action bits and ALU operations per stage. This often requires an {\em expert} with intimate knowledge of the product.

\begin{table}[t]
    \centering
    \small
    \setlength\extrarowheight{1.25pt}
    \setlength\tabcolsep{2.5pt}
    \begin{tabular}{lcccc}
    \hline
      \thead[l]{Scheme} & \thead[c]{TCAM \\ Blocks} & \thead[c]{SRAM \\ Pages} & \thead[c]{Steps \\ (Stages)} & \thead[c]{Model} \\ 
        \hline
        \resail ($min\_bmp$=13) & 1.14 & 549.12 & 2 & CRAM \\
        \resail ($min\_bmp$=13) & 2 & 556 & 9 & Ideal RMT \\
        \resail ($min\_bmp$=13) & 17 & 750 & 16 & Tofino-2 \\
        \bottomrule
    \end{tabular}
    \vspace{-0.5em}
    \caption{Predictive accuracy of \cram for \resail (IPv4)}
    \label{tab:ipv4_cram_evaluation}
    \vspace{-0.5em}
\end{table}

\begin{table}[t]
    \centering
    \small
    \setlength\extrarowheight{1.25pt}
    \setlength\tabcolsep{2.5pt}
    \begin{tabular}{lcccc}
    \hline
      \thead[l]{Scheme} & \thead[c]{TCAM \\ Blocks} & \thead[c]{SRAM \\ Pages} & \thead[c]{Steps \\ (Stages)} & \thead[c]{Model} \\ 
        \hline
        \bsic ($k$=24) & 7.45 & 203.52 & 14 & CRAM \\
        \bsic ($k$=24) & 15 & 211 & 14 & Ideal RMT \\
        \bsic ($k$=24) & 15 & 416 & 30 & Tofino-2 \\
        \bottomrule
    \end{tabular}
    \vspace{-0.5em}
    \caption{Predictive accuracy of \cram for \bsic (IPv6)}
    \label{tab:ipv6_cram_evaluation}
    \vspace{-\baselineskip}
\end{table}

Consider the predictive accuracy for \resail in \cref{tab:ipv4_cram_evaluation} as we move from \cram to ideal RMT. The TCAM and SRAM measures reflect small rounding errors due to unit conversion. However, the latency increases significantly from 2 steps to 9 stages because, unlike dRMT, RMT stages provide {\em both} memory and processing---to support 556 RAM pages, more stages are required even when no additional processing is needed.

Next, consider the predictive accuracy for \resail in \cref{tab:ipv4_cram_evaluation} as we move from ideal RMT to Tofino-2. There is a small additive increase in TCAM blocks due to extra ternary tables required for implementing \resail in P4. Additionally, SRAM pages increase by a factor of 1.35, and stages increase by a factor of 1.78. As discussed earlier in \cref{subsubsec:ipv4_comparison}, this is because the maximum achievable SRAM utilization on Tofino-2 is 50\%, necessitating more SRAM pages and stages.

The predictive accuracy for \bsic in \cref{tab:ipv6_cram_evaluation} can be interpreted similarly. The key difference is that the $\sim$2X increase in SRAM pages and stages from ideal RMT to Tofino-2 is due to the fact that implementing 3-way branching on Tofino-2 requires two stages for each BST level.

Based on our limited experience, the \cram model provides a useful, easily computed initial model for estimating algorithm scalability. Although its measures of space and time are off by small constant factors, this is no different from Big O notation.
\section{Related Work}
\label{sec:related}

\textbf{Models of Computation:} Abstract models like RAM~\cite{ram}, PRAM~\cite{pram}, and Turing machines~\cite{models_of_computation} are widely used to analyze algorithms. The RAM model abstracts sequential computers, while the PRAM model abstracts shared memory multiprocessors. Our \cram model abstracts network processors with two types of memory, parallelism, and programmability.

\textbf{Combinations of CAM and RAM:} Earlier CAM and RAM combinations optimize different metrics. CoolCAM~\cite{coolcam}, cooler TCAM~\cite{coolertcam}, and EaseCAM~\cite{easecam} all reduce power consumption. Liu~\cite{liu} compacts routing tables to reduce cost, power consumption, and thermal dissipation. Luo~\cite{virtual_routers} merges FIBs in virtual routers to reduce TCAM usage. Compaction and merging are orthogonal to our ideas. Other hybrid approaches~\cite{magictcam, extendedtcam, portcatcher, treecam} target tasks such as packet classification. In summary, no existing solution optimizes scalability for IP lookup by leveraging {\em both} TCAM and SRAM.
\section{Conclusion}
\label{sec:conclusion}

Our paper introduces new algorithms for IP lookup made possible by strategically leveraging {\em both} CAM and RAM. These algorithms address the scaling challenges of global routing tables and are well-suited to the architectures of modern network processors. For these processors, the \cram lens provided a fresh perspective on algorithm design. Much like the RAM~\cite{ram} and PRAM~\cite{pram} models, \cram offers metrics for quickly evaluating algorithm feasibility {\em before} implementation. Using the \cram model, we developed three new IP lookup algorithms: \resail, \bsic, and \mashup. \resail and \bsic scale to much larger databases than the best existing IPv4 and IPv6 lookup schemes, respectively, while \mashup excels in stage-constrained hardware environments.

We aim to establish \cram's generality by applying it to other hardware architectures~(\cref{subsec:other_architectures}) and network applications~(\cref{subsec:other_applications})---helping to ``cram'' more packet processing power into each unit of chip area. Ultimately, our findings underscore a simple insight for networking chip vendors: \textit{a little TCAM goes a long way. Adding small amounts of TCAM to supplement SRAM can significantly improve scalability.}


\section*{Acknowledgments}

We thank our shepherd, Xiaoqi Chen, and the anonymous reviewers for their valuable feedback. We are also grateful to David Maltz and Vladimir Gurevich for their guidance. This research was supported in part by NSF grant CNS-2333587.


\bibliographystyle{plain}
\bibliography{reference}

\clearpage
\appendix

\section{Appendix}
\label{sec:appendix}

\subsection{Packet Processing Architectures} \label{subsec:rmt_drmt_overview}

\begin{figure}[h]
    \vspace{-1em}
     \centering
     \begin{subfigure}[h]{.42\textwidth}
         \centering
        \includegraphics[width=\textwidth]{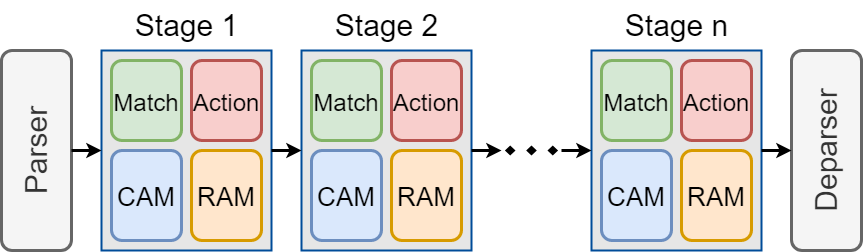}
        \caption{RMT architecture}
         \label{fig:rmt}
     \end{subfigure}
     \begin{subfigure}[h]{.42\textwidth}
         \centering
        \includegraphics[width=\textwidth]{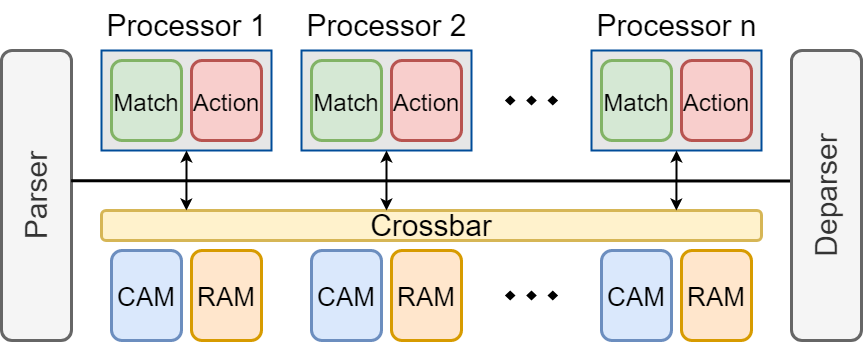}
         \caption{dRMT architecture}
         \label{fig:drmt}
     \end{subfigure}
     \caption{RMT vs dRMT}
    \label{fig:rmt_drmt}
\end{figure}

\vspace{-\baselineskip}
\vspace{-0.6em}

\subsection{RMT and dRMT Implementations} \label{subsec:rmt_drmt_implementations}
\vspace{-0.5em}

\begin{table}[h]
    \centering
    \small
    \setlength\extrarowheight{1.25pt}
    \setlength\tabcolsep{2.5pt}
    \begin{tabular}{lccc}
    \hline
      \thead[l]{Product} & \thead[c]{Architecture} & \thead[c]{Type} & \thead[c]{Sources} \\ 
        \hline
        Intel Tofino-2 & RMT & switch ASIC & \cite{tofino_rmt, tofino_slides, intel_tofino} \\
        Intel Mount Evans (E2000) & RMT & SmartNIC & \cite{e2000, intel_ipu} \\
        AMD Pensando DSC-100 & RMT & SmartNIC & \cite{amd_pensando, pensando_arch, pensando_p4} \\
        Fungible F1/S1 & RMT & SmartNIC & \cite{fungible} \\
        FlowBlaze & RMT & FPGA & \cite{flowblaze} \\
        FlexCore & dRMT & switch ASIC & \cite{flexcore} \\
        Nvidia BlueField-3 & dRMT & SmartNIC & \cite{nvidia_bluefield, nvidia_smartnic, nvidia_smartnic_2} \\
        \bottomrule
    \end{tabular}
    \caption{Summary of known implementations}
    \label{tab:implementations}
\end{table}

While we cannot verify all details in some cases, it is evident that TCAM, SRAM, parallelism, and programmability are present in the sources for the products listed in \cref{tab:implementations}.

\vspace{-0.5em}

\subsection{Updates, Deletions, and Insertions} \label{subsec:updates}

\subsubsection{\resail} \label{subsubsec:resailupdates}

Incremental updates, deletions, and insertions for \resail are efficient, following the same process as lookups but with modifications to the target entry. For prefixes of length $min\_bmp$ or greater, only two memory accesses are required~(bitmap and hash table). For prefixes shorter than $min\_bmp$, the operations are more costly because of prefix expansion. Update operations for prefixes longer than 24 bits are much simpler in \resail than in \sail because we have eliminated pivot pushing.

\vspace{-0.5em}

\subsubsection{\bsic} \label{subsubsec:bsicupdates}

For \bsic, incremental updates, deletions, and insertions are costly and complex due to inherent dependencies between binary search tree levels. A separate database with additional prefix information is needed for rebuilding data structures~\cite{dxr}. If fast update operations are important, \resail and \mashup are better choices.

\vspace{-0.5em}

\subsubsection{\mashup} \label{subsubsec:mashupupdates}
Incremental updates, deletions, and insertions for \mashup are nearly identical to lookups, except they modify the target entry. These are standard algorithms~\cite{network_algorithmics} for multibit tries that have been well studied. Maintaining a sorted TCAM table under these changes is non-trivial, but effective algorithms exist~\cite{tcam_updates}.

\vspace{-0.5em}

\subsection{Range Expansion and Optimizations} \label{subsec:bsic_details}

\vspace{-0.5em}

\begin{table}[h]
    \centering
    \small
    \setlength\extrarowheight{1.25pt}
    \setlength\tabcolsep{2.5pt}
    \begin{tabular}{|c|c|}
    \hline
    \thead[c]{Range} & \thead[c]{Next Hop} \\ 
    \hline
    0000 - 0011 & C \\
    0100 - 0100 & A \\
    0101 - 0111 & D \\
    1000 - 1001 & - \\
    1010 - 1010 & B \\
    1011 - 1011 & C \\
    1100 - 1111 & - \\
    \bottomrule
    \end{tabular}
    \caption{Range expansion for slice 1001 (\cref{tab:initial_lookup_table})}
    \label{tab:expansion_example}
\end{table}

\vspace{-\baselineskip}

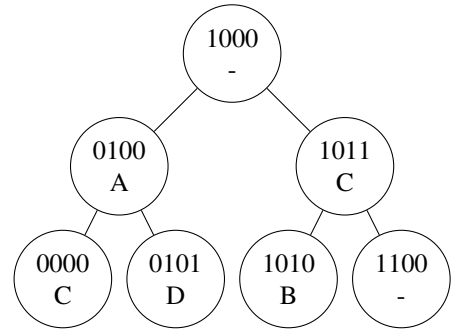
\begin{figure}[h]
    \centering
    \begin{tikzpicture}[
      every node/.style = {minimum width = 2em, draw, circle, align=center},
      level/.style = {sibling distance = 30mm/#1}
      ]
      \node {1000\\-}
      child {node {0100\\A} 
            child {node {0000\\C}}
            child {node {0101\\D}}
            }
      child {node {1011\\C}
            child {node {1010\\B}}
            child {node {1100\\-}}
            };
    \end{tikzpicture}
    \caption{BST for slice 1001 (\cref{tab:initial_lookup_table})}
    \label{fig:tree_example}
\end{figure}

As in \dxr~\cite{dxr}, convert all the prefix substrings into ranges by generating their endpoint pairs. Use the endpoint pairs to create sorted, contiguous, and non-overlapping intervals that cover all possible bitstrings of the maximum length. Intervals that are added to complete the full range will "inherit" the next hop of the current lookup table entry's longest prefix match. This is necessary because it is possible for a destination address to be incorrectly directed by the initial lookup table to a BST that does not contain a legitimate match. Therefore, in the case of such a mistake, the search key will land in an interval containing the correct next hop. A simple example of range expansion for slice 1001 from \cref{tab:initial_lookup_table} is shown in \cref{tab:expansion_example}. Note that the intervals 1000-1001 and 1100-1111 were added to complete the full range. Since there are no valid longest prefix matches for slice 1001, its intervals are assigned a default value of -.

After the full range is created, merge neighboring intervals with the same next hop to reduce the number of nodes required. Discard the right endpoints as they can be inferred from the left ones. Use the remaining left endpoints to create the BST. \cref{fig:tree_example} shows a sample BST for slice 1001.

\vspace{-1em}

\subsection{Pseudocode} \label{subsec:pseudocode}

\begin{algorithm}[h]
    \SetKw{Break}{break}
    \SetKwInOut{KwIn}{Input}
    \SetKwInOut{KwOut}{Output}
    \KwIn{$addr$, IPv4 address}
    \KwIn{$min\_bmp$, smallest bitmap available}
    \KwOut{$hop$, next hop}
    $hop$ $\leftarrow$ $lookup\_table$.match($addr$)

    \If{$hop$ $\neq$ None}{\KwRet{$hop$}}
    
    \For{$i$ $\leftarrow$ 24 ; $i$ $\geq$ $min\_bmp$ ; $i$\texttt{{-}{-}}}{
        \If{$B_{i}$.match($addr$ $\gg$ (32-$i$)) \texttt{==} 1}
        {
            $key$ $\leftarrow$ ($addr$ $\gg$ (32-$i$)) $\ll$ (25-$i$)

            $key$ $\leftarrow$ $key$ + (1 $\ll$ (24-$i$))
            
            $hop$ $\leftarrow$ $hash\_table$.match($key$)

            \Break
        }
    }
    \KwRet{$hop$}
    \caption{\textbf{\resail Lookup} ($addr$, $min\_bmp$)} \label{alg:resail_lookup}
\end{algorithm}

\begin{algorithm}[h]
    \SetKw{Break}{break}
    \SetKwInOut{KwIn}{Input}
    \SetKwInOut{KwOut}{Output}
    \KwIn{$addr$, IPv4 or IPv6 address}
    \KwIn{$k$, initial slice size}
    \KwOut{$hop_{best}$, next hop}
    \SetKwFor{While}{while}{}{end while}%
    $level$ $\leftarrow$ 0

    $len$ $\leftarrow$ 32 if IPv4, 64 if IPv6
    
    $hop_{best}$,$index$ $\leftarrow$ $lookup\_table$.match($addr$ $\gg$ ($len-k$))

    \If{$hop_{best}$ $\neq$ None}{\KwRet{$hop_{best}$}}

    \While{$index$ $\neq$ None}{
        $hop$, $left$, $right$, $prefix$ $\leftarrow$ $bst_{level}$.match($index$)

        \If{$prefix$ \texttt{==} (($addr$ $\ll$ $k$) $\gg$ $k$)}{
            \KwRet{$hop$}
        }
        \ElseIf{$prefix$ < (($addr$ $\ll$ $k$) $\gg$ $k$)}{
            $index$ $\leftarrow$ $right$
            
            $hop_{best}$ $\leftarrow$ $hop$
        }
        \Else{
            $index$ $\leftarrow$ $left$
        }
        $level$ $\leftarrow$ $level$ + 1
    }
    \KwRet{$hop_{best}$}
    \caption{\textbf{\bsic Lookup} ($addr$, $k$)} \label{alg:bsic_lookup}
\end{algorithm}

\begin{algorithm}[h]
    \SetKw{Break}{break}
    \SetKwInOut{KwIn}{Input}
    \SetKwInOut{KwOut}{Output}
    \KwIn{$addr$, IPv4 or IPv6 address}
    \KwIn{$strides$, set of strides}
    \KwOut{$hop_{best}$, next hop}
    \SetKwFor{While}{while}{}{end while}%

    $hop_{best}$ $\leftarrow$ $default$
    
    $table$ $\leftarrow$ $root$

    $tag$ $\leftarrow$ None
    
    $index$ $\leftarrow$ 0

    $level$ $\leftarrow$ 0

    \While{$table$ $\neq$ None}{
        $key$ $\leftarrow$ $addr$[$index$:$index$+$strides$[$level$]]
        
        $hop$, $next\_table$, $tag$ $\leftarrow$ $table$.match($tag$, $key$)

        \If{$table$.hit()}{
            \If{$hop$ $\neq$ None}{
                $hop_{best}$ $\leftarrow$ $hop$
            }
            
            $index$ $\leftarrow$ $index$ + $strides$[$level$]
            
            $level$ $\leftarrow$ $level$ + 1

            $table$ $\leftarrow$ $next\_table$
        }
        \Else{
            \Break
        }
    }
    \KwRet{$hop_{best}$}
    \caption{\textbf{\mashup Lookup} ($addr$, $strides$)} \label{alg:mashup_lookup}
\end{algorithm}


\subsection{Latency-memory trade-offs} \label{subsec:tradeoff}

A natural question is: {\em are there latency-memory trade-offs for CRAM algorithms that we can exploit to free up pipeline stages for other processing tasks?} Examining the basic \cram model, there appears to be a clear trade-off: reducing dependency steps at the expense of increased memory or computation can lead to lower latency. 

Unfortunately, on real platforms like Tofino-2, {\em steps} cannot be conflated with {\em stages} because, in RMT, stages provide a fixed amount of both memory and computation. Therefore, large lookup tables require multiple stages. While a latency-memory trade-off exists for {\em steps} versus memory in our three \cram algorithms, no corresponding trade-off exists for {\em stages} versus memory. Instead, there is an optimal number of stages, beyond which {\em both} memory and latency increase.

\begin{figure}[h]
    \centering
    \begin{tikzpicture}
    \large
    \begin{axis}[
        xmin = 12, xmax = 44,
        ymin = 0, ymax = 100,
        xtick={12,16,...,44},
        ytick={0,10,...,100},
        width=\columnwidth*0.9,
        height=170pt,
        xticklabel={\ifdim\tick pt<10pt 0\fi\pgfmathprintnumber{\tick}},
        xticklabel style={rotate=0,font=\small},
        yticklabel style={rotate=0,font=\small},
        xlabel style={font=\normalsize,at={(axis description cs:.5,-.08)}},
        ylabel style={font=\normalsize,at={(axis description cs:-.07,.5)}},
        legend columns=3,
        legend style={at={(-0.12,-0.22)},anchor=north west,font=\small},
        xlabel=Slice Size $k$,
        ylabel=\% of Tofino-2 Capacity,
        tick pos=left
    ]
    \addplot[line width=1.5pt,mark=none,smooth,color=blue]
        coordinates{
            (12,0.2083333333)
            (16,0.2083333333)
            (20,0.4166666667)
            (24,3.125)
            (28,8.125)
            (32,15.20833333)
            (36,25)
            (40,35.20833333)
            (44,48.75)
        };
    \addlegendentry{TCAM blocks}
    \addplot[line width=1.5pt,mark=none,densely dashed,color=red]
        coordinates{
            (12,14.6875)
            (16,13.3125)
            (20,13.125)
            (24,13.1875)
            (28,12.125)
            (32,9.6875)
            (36,8.6875)
            (40,7.25)
            (44,5.9375)
        };
    \addlegendentry{SRAM pages}

    \addplot[line width=1.5pt,mark=none,densely dotted,color=green]
        coordinates{
            (12,90)
            (16,85)
            (20,80)
            (24,70)
            (28,75)
            (32,85)
            (36,85)
            (40,90)
            (44,100)
        };
    \addlegendentry{Stages}
    
    \end{axis}
    \end{tikzpicture}
    \caption{\bsic IPv6 latency-memory trade-off on an ideal RMT chip for AS131072 (Sep 2023)}
    \label{fig:latency_memory_tradeoff}
\end{figure}
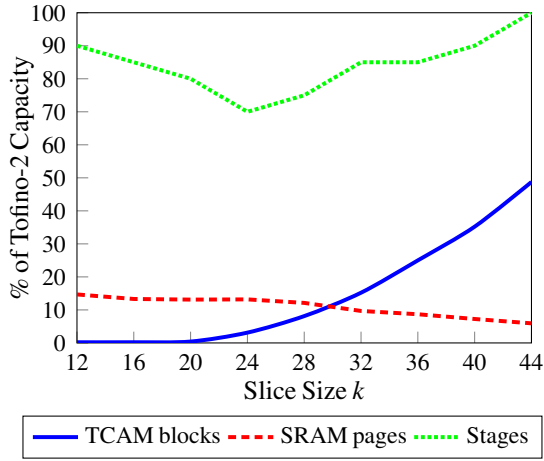

To see this, consider \cref{fig:latency_memory_tradeoff} for \bsic~(IPv6), in which the only tuning parameter is $k$---the width of the initial TCAM table. While a latency-memory curve exists, decreasing latency by increasing $k$ actually {\em increases} the number of stages required. As $k$ grows, the number of stages needed for the initial TCAM table rises significantly, outweighing the reduction in stages gained from decreasing BST depth. In contrast, the basic \cram model predicts reduced latency as $k$ increases. Unfortunately, this larger TCAM table requires more {\em stages} but not {\em steps}.

As shown in \cref{fig:latency_memory_tradeoff}, the optimal value of $k$ is 24, with both smaller and larger values yielding worse results. Thus, no trade-off exists between stages and memory. This is also the reason we use $k=24$ in our experiments for IPv6.

Similarly, with \mashup, the strides are the primary tuning parameter. Again, we did not find a useful memory-latency trade-off when considering stages. Finally, \resail consistently requires only two steps, with no latency-memory trade-off whatsoever in either the \cram or ideal RMT models.

While there is no useful latency-memory trade-off for a {\em fixed} database size in our three algorithms, an obvious trade-off emerges as the database size {\em increases}: the number of stages~(latency) must increase, at least to provide more memory. This is implicit in the linear trade-off curves shown in \cref{fig:ipv4_scaling} and \cref{fig:ipv6_scaling}. Notably, the y-axes in these figures represent SRAM pages. However, since more SRAM requires proportionally more stages, \cref{fig:ipv4_scaling} and \cref{fig:ipv6_scaling} can be rescaled to~(instead) depict stages versus database size, maintaining exactly the same curve shapes.

\end{document}